\documentclass[prd,aps,showpacs,nofootinbib,preprint,eqsecnum]{revtex4}
\usepackage{graphicx,color,amsmath,amsxtra}
\usepackage{amssymb}
\usepackage[english]{babel}
\usepackage{amsfonts}
\usepackage{ulem}
\allowdisplaybreaks[4]

\newcommand{\beq}{\begin{equation}}
\newcommand{\eeq}{\end{equation}}
\newcommand{\bea}{\begin{eqnarray}}
\newcommand{\eea}{\end{eqnarray}}
\newcommand{\bseq}{\begin{subequations}}
	\newcommand{\eseq}{\end{subequations}}

\newcommand{\Ref}[1]{(\ref{#1})}

\begin{document}

\title{Cosmology with nonminimal kinetic coupling and a Higgs-like potential}

\author{Jiro Matsumoto\footnote{E-mail address: jmatsumoto@kpfu.ru}
and Sergey V.~Sushkov\footnote{E-mail address: sergey\_sushkov@mail.ru}}
\affiliation{Institute of Physics, Kazan Federal University, Kremlevskaya Street 18,
Kazan 420008, Russia}
	
\begin{abstract}
We consider cosmological dynamics in the theory of gravity with the scalar field possessing the nonminimal kinetic coupling to curvature given as $\kappa G^{\mu\nu}\phi_{,\mu}\phi_{,\nu}$, and the Higgs-like potential $V(\phi)=\frac{\lambda}{4}(\phi^2-\phi_0^2)^2$.
Using the dynamical system method, we analyze stationary points, their stability, and all possible asymptotical regimes of the model under consideration.
We show that the Higgs field with the kinetic coupling provides an existence of accelerated regimes of the Universe evolution. There are three possible cosmological scenarios with  acceleration:
(i) {\it The late-time  de Sitter epoch} when the Hubble parameter tends to the constant value, $H(t)\to H_\infty=(\frac23 \pi G\lambda\phi_0^4)^{1/2}$ as $t\to\infty$, while the scalar field tends to zero, $\phi(t)\to 0$, so that the Higgs potential reaches its local maximum $V(0)=\frac14 \lambda\phi_0^4$.
(ii) {\it The Big Rip} when $H(t)\sim(t_*-t)^{-1}\to\infty$ and $\phi(t)\sim(t_*-t)^{-2}\to\infty$ as $t\to t_*$.
(iii) {\it The Little Rip} when $H(t)\sim t^{1/2}\to\infty$ and $\phi(t)\sim t^{1/4}\to\infty$ as $t\to\infty$. Also, we derive modified slow-roll conditions for the Higgs field and demonstrate that they lead to the Little Rip scenario.
\end{abstract}
	
	
\maketitle


\section*{Introduction}
Various precise observations in astronomy, e.g.
the observations of the Ia type supernovae (SNIa) \cite{supernova},
the cosmic microwave background radiation (CMB)\cite{CMB},
the baryon acoustic oscillations (BAO)\cite{BAO}, and
the large scale structure of the Universe (LSS),
clarified that the current Universe
is acceleratedly expanding. 
This acceleration cannot be explained by usual matter, and therefore so-called dark energy, which has a negative pressure with an equation of state parameter $w<-1/3$, is introduced.
The most famous candidate for dark energy is the cosmological constant $\Lambda$,
and the $\Lambda$CDM model, which assumes the existence of the cosmological constant $\Lambda$ and the cold dark matter (CDM), represents the standard model in cosmology.
In the $\Lambda$CDM model a contribution of the spatial curvature into the total energy balance  gives is $\rho _K \propto a^{-2}$. Also, it is known that the current energy fraction of the spatial curvature $\Omega _{K0}$ is negligible. This fact leads to the condition $\Omega _{K} \ll \Omega _{m}$ in the early time of the Universe, because the energy density of nonrelativistic matter is proportional to $a^{-3}$. This is known as the flatness problem of the Universe which, in turn, entails the problem of fine-tuned initial conditions. To overcome this and a number of other problems, the concept of cosmological inflation, i.e. accelerated expansion of the Universe in very early cosmic times, has been introduced in theoretical physics.

The challenge for theoretical cosmology is an adequate description of the whole history of the Universe evolution including two epochs of accelerated expansion and the matter-dominated phase. To solve this problem, at least partially, there exists a great many of models
mostly based on phenomenological ideas which involve new dynamical sources of gravity that act as dark energy, and/or various modifications to general relativity.
In particular, quintessence \cite{quintessence}, ghost condensate \cite{Ghost}, and $k$-essence models \cite{Kessense} assume the existence of a dynamical scalar field spreading over the whole Universe instead of the cosmological constant. On the other hand,  scalar-tensor theories of gravity \cite{jordan,bd,Fujii_Maeda} and $F(R)$ gravity \cite{fr,Nojiri:2006ri,Sotiriou:2008rp,DeFelice:2010aj,Nojiri:2010wj} represent modified theories of gravity concerning the accelerated expansion of the Universe.

In this paper, we consider a scalar field model which has a nonminimal derivative coupling with gravity. It is known that such the model can give rise to de Sitter expansion without potential terms of the scalar field \cite{Sus:2009}.
Therefore, it could be a candidate of dark energy or the cause of inflation.
Further investigations of cosmological and
astrophysical models with nonminimal derivative couplings have been continued in
\cite{kincoupl,Bruneton,SarSus:2010,Sus:2012,SkuSusTop:2013,SusRom:2012}.
Note that generally the order of field equations in models with nonminimal
derivative couplings is higher than two. However, it reduces to second order in
the particular case when the kinetic term is only coupled to the Einstein
tensor, i.e. $\kappa G_{\mu\nu}\phi^{,\mu}\phi^{,\nu}$ (see, for example, Ref.
\cite{Sus:2009}).\footnote{It is worth noting that a general single scalar field
	Lagrangian giving rise to second-order field equations had been derived by Horndeski
	\cite{Horndeski} in 1974. The model with $\kappa G_{\mu\nu}\phi^{,\mu}\phi^{,\nu}$
	represents a particular form of the Horndeski Lagrangian. Recent interest in second-order
	gravitational theories is also connected with the Dvali-Gabadadze-Porrati
	braneworld \cite{DGP} and and Galileon gravity \cite{Ggravity}.}
We will consider a Higgs-like potential of the scalar field in this paper.
This kind of model is investigated in the inflation model called new Higgs inflation \cite{Germani:2010gm}, however, the dynamics of the scalar field has not been well investigated because the slow-roll approximation is assumed in advance.
In this paper the dynamics of the scalar field is considered by regarding the field equation as an autonomous system without using approximations.

The paper is organized as follows. In Sec.~\ref{sec2}, we discuss a general action and field equations in the theory of gravity with a scalar field possessing the nonminimal kinetic coupling to the curvature. In Sec.~\ref{sec3}, we apply the standard method of autonomous dynamical systems to investigate dynamics of the scalar field with a Higgs-like potential. The particular attention is given to  an asymptotical behavior of the scalar field.
Expansion scenarios of the Universe corresponding to the asymptotics found in Sec.~\ref{sec3} are investigated in Sec.~\ref{CosS}.
Dynamics of the Universe in new Higgs inflation model is also discussed.
The case that the scalar field is regarded as Higgs field is treated in Sec.~\ref{Higgs}.
Concluding remarks are given in Sec.~\ref{Conclusion}.


\section{Action and field equations \label{sec2}}
Let us consider the theory of gravity with the action
\begin{equation}\label{action}
S=\int d^4x\sqrt{-g}\left\{ \frac{R}{8\pi G} -\big[g^{\mu\nu} + \kappa G^{\mu\nu}
\big] \phi_{,\mu}\phi_{,\nu} -2V(\phi)\right\},
\end{equation}
where $V(\phi)$ is a scalar field potential, $g_{\mu\nu}$ is a metric, $R$ is the
scalar curvature, $G_{\mu\nu}$ is the Einstein tensor.
The coupling parameter $\kappa$ has the dimension of inverse mass-squared.
Note that in the literature there is discussion, which is still open, about acceptable values of $\kappa$. In Refs. \cite{Germani:2010gm,GermaniKehagias:2011} 
it was assumed that $\kappa<0$ in order to prevent the appearance of ghosts in the model. 
However, analyzing scalar and tensor perturbations generated in the theories given by the action \Ref{action}, 
Tsujikawa in Ref. \cite{Tsujikawa} had derived more general conditions in order to avoid the appearance of scalar ghosts and Laplacian instabilities. 
Generally, their fulfillment depends on particular cosmological scenarios and is not directly determined by the sign of $\kappa$. 
Moreover, in Ref. \cite{Dent} it was demonstrated that the necessary condition that ensures the absence of instabilities is 
fulfilled for $\kappa$ both positive and negative. Since the detailed investigation of the instabilities and superluminality of 
the model with nonminimal kinetic coupling lies beyond the scope of the present work, hereinafter we will not impose any restrictions on values of $\kappa$.

In the spatially-flat Friedmann-Robertson-Walker cosmological model the action
\Ref{action} yields the following field equations \cite{Sus:2012}
\bseq\label{genfieldeq}
\bea
\label{00cmpt}
&&3H^2=4\pi G\dot{\phi}^2\left(1-9\kappa H^2\right) +8\pi G V(\phi),\\
&&\displaystyle
2\dot{H}+3H^2=-4\pi G\dot{\phi}^2
\left[1+\kappa\left(2\dot{H}+3H^2 +4H\ddot{\phi}\dot{\phi}^{-1}\right)\right]
+8\pi G V(\phi),
\label{11cmpt}
\\
\label{eqmocosm}
&&(\ddot\phi+3H\dot\phi)-3\kappa(H^2\ddot\phi
+2H\dot{H}\dot\phi+3H^3\dot\phi)=-V_\phi,
\eea
\eseq
where a dot denotes derivatives with respect to time, $H(t)=\dot a(t)/a(t)$ is
the Hubble parameter, $a(t)$ is the scale factor, $\phi(t)$ is a homogeneous
scalar field, and $V_\phi=dV/d\phi$.
Note that equations \Ref{11cmpt} and \Ref{eqmocosm} are of second
order, while \Ref{00cmpt} is a first-order differential
constraint for $a(t)$ and $\phi(t)$. The constraint (\ref{00cmpt}) can be
rewritten as
\beq\label{constrphigen}
\dot\phi^2=\frac{3H^2-8\pi G
	V(\phi)}{4\pi G(1-9\kappa H^2)},
\eeq
or equivalently as
\beq\label{constralphagen}
H^2=\frac{4\pi G\dot\phi^2+8\pi G
	V(\phi)}{3(1+12\pi G\kappa\dot\phi^2)}.
\eeq
Therefore, as long as the parameter $\kappa$ and the potential
$V(\phi)$ are given, the above relations provide restrictions for the
possible values of $H$ and $\dot\phi$, since they have to give rise to
non-negative $\dot\phi^2$ and $H^2$, respectively. Assuming the non-negativity of the potential, i.e. $V(\phi)\ge0$, we can conclude from Eqs. \Ref{constrphigen} and \Ref{constralphagen} that in the theory with the positive $\kappa$ possible values of $\dot\phi$ are unbounded, while $H$ takes restricted values. Vice versa, the negative $\kappa$ leads to bounded $\dot\phi$ and unbounded $H$.

Let us now resolve the equations \Ref{11cmpt} and \Ref{eqmocosm} with
respect to $\dot H$ and $\ddot\phi$, and then, using the relations
\Ref{constrphigen} and \Ref{constralphagen}, eliminate
$\dot\phi$ and $H$ from respective equations. As the result, we obtain
\begin{equation}\label{a2gen}
\dot H =\frac{-(1-3\kappa H^2)(1-9\kappa H^2)[3H^2-8\pi G V(\phi)]-
	4\sqrt{\pi G}\kappa H \sqrt{(1-9\kappa H^2)[3H^2-8\pi G V(\phi)]}\,V_\phi}
{1-9\kappa H^2+54\kappa^2 H^4-8\pi G\kappa V(\phi)(1
	+9\kappa H^2)},
\end{equation}
\begin{align}\label{phi2gen}
\ddot\phi=& \big[1+12\pi G \kappa\dot\phi^2+96\pi^2 G^2 \kappa^2\dot\phi^4
	+8\pi G \kappa V(\phi)(12\pi G \kappa\dot\phi^2-1)\big]^{-1}
        \nonumber \\
        & \times \bigg \{ -2\sqrt{3\pi G}\dot\phi
	[1+8\pi G \kappa\dot\phi^2-8\pi G \kappa V(\phi)]
	\sqrt{[\dot{\phi}^2+2V(\phi)](12\pi G \kappa\dot\phi^2+1)}  \nonumber \\
        &-(12\pi G \kappa\dot\phi^2+1)(4\pi G \kappa\dot\phi^2+1)V_\phi \bigg \} .
\end{align}
It is seen that the $H$-equation \Ref{a2gen} in general contains
$\phi$-terms arising from the potential $V(\phi)$. At the same time, the $\phi$-equation \Ref{phi2gen} does not contains $H$-terms, so that it represents a closed second-order differential equation for the unknown function $\phi(t)$. Such the form of Eq. \Ref{phi2gen} dictates us the following strategy of solving the system of field equations \Ref{genfieldeq}. First, analyzing the $\phi$-equation \Ref{phi2gen}, we will characterize in detail a time evolution of the scalar field $\phi(t)$. Then, using the constraint \Ref{constralphagen}, we will be able to describe an evolution of the Hubble parameter $H(t)$.

\section{Scalar field as an autonomous dynamical system \label{sec3}}
In this section we will focus on solving Eq. \Ref{phi2gen} which describes a time evolution of the scalar field. Hereafter we specify the scalar potential in the Higgs-like form:
\begin{equation}
V(\phi)=\frac{\lambda}{4} (\phi ^2 - \phi _0^2)^2,
\label{Higgspotential}
\end{equation}
where $\lambda$ and $\phi _0$ are positive constants.

Note that Eq. \Ref{phi2gen} has a normal form, i.e. resolved with respect to the second derivative.
This allows us to use standard methods of the theory of dynamical systems.
In practice, an analysis of Eq. \Ref{phi2gen} depends on particular values of the coupling parameter $\kappa$,
and hence further we will separately consider the cases $\kappa>0$, $\kappa<0$, and $\kappa=0$.

\subsection*{The case $\kappa > 0$}
Assume that $\kappa$ is positive.
Now, let us introduce the set of dimensionless variables
\beq
x = \frac{\phi}{\phi_0},\quad
y = \sqrt{8 \pi G\kappa}\, \dot \phi, \quad
\tau = \phi _0 t,
\eeq
and parameters
\beq
V_0 = 2 \pi G\kappa \lambda \phi _0 ^4,\quad
\gamma = G \phi _0 ^2.
\eeq
Substituting $V(\phi)$ given by Eq. \Ref{Higgspotential} and using dimensionless variables, we can rewrite Eq. \Ref{phi2gen} as the following autonomous dynamical system:
\begin{align}
\frac{dx}{d \tau} &= 
\sqrt{\frac{\lambda}{4V_0}}\, y,
\label{A1} \\
\frac{dy}{d \tau} &
= \frac{1}{\Delta}
 \left \{ - \sqrt{3 \pi \gamma \lambda V_0^{-1}}\, y \left [ 1+y^2-V_0 (x^2-1)^2 \right ] \sqrt{[y^2+2V_0(x^2-1)^2]\left({\textstyle\frac{3}{2}}y^2+1 \right)}
\right. \nonumber \\
& \textstyle
- 2 \sqrt{\lambda V_0} \left ( \frac{3}{2}y^2 +1 \right ) \left ( \frac{1}{2}y^2 +1 \right ) x(x^2-1) \bigg \}, \label{A2}
\end{align}
where
$$\textstyle
\Delta={1+\frac{3}{2}y^2+\frac{3}{2}y^4+V_0 (x^2-1)^2 \left ( \frac{3}{2}y^2 -1 \right )}.
$$

Below we examine in detail basic features of the system (\ref{A1})-(\ref{A2}).

\subsection{Stationary points}
Stationary points are those where $dx/d\tau=0$ and $dy/d\tau=0$. 
Hence, from Eq. \Ref{A1} it follows that any stationary point of the system (\ref{A1})-(\ref{A2}) has $y=0$. Then, Eq. \Ref{A2} yields
\beq
\left.\frac{dy}{d \tau}\right|_{y=0} =\frac{- 2 \sqrt{\lambda V_0} x(x^2-1)}{1-V_0 (x^2-1)^2}.
\eeq
The equality $(dy/d\tau)_{y=0}=0$ is fulfilled if $x=0$ for $V_0\not=1$, and $x=\pm1$ for any $V_0$. 
Therefore, $(x,y)=(0,0)$ is the stationary point of the system (\ref{A1})-(\ref{A2}) if $V_0\not=1$, 
and $(x,y)=(\pm 1, 0)$ are the stationary points for any $V_0$.
Also, it is necessary to stress that for any $V_0$ the value $(dy/d\tau)_{y=0}$ obeys the limit $(dy/d\tau)_{y=0}\to 0$ as $x\to\pm\infty$. 
This means that $(x,y)=(\pm\infty, 0)$ are also the stationary points. 
Thus, the complete list of stationary points of the system (\ref{A1})-(\ref{A2}) is 
represented as $(x,y)=(0,0)$, $(\pm 1, 0)$, and $(\pm\infty,0)$.\footnote{It is worth noticing that one should be careful with treating critical points at infinity. Strictly speaking, if the phase space is non-compact, one must use the Poincar\'{e} central projection method \cite{Poincare1,Poincare2,Poincare3,Poincare4,Poincare5} to investigate the dynamics at infinity. The use of this method is especially necessary if critical points are lying on some `infinite' $n$-dimensional surface with $n\le 2$. The essence of the method consists in using the specific Poincar\'{e} coordinates which convert the surface lying at infinity into a compact surface. 
We thank an anonymous referee for drawing our attention to this problem.  However, in our case the situation is more simple. We have only two dynamical variables, $x$ and $y$, and any critical point has $y=0$. So, it is only necessary to check either the `infinite' points with $x=\pm\infty$ are critical or not. This could be easily done directly, without applying the Poincar\'{e} central projection method.}

The stability of stationary points with respect to small perturbations $\delta x$ and $\delta y$ can be investigated by fluctuating Eqs.~(\ref{A1})-(\ref{A2}) as
\begin{equation}
\frac{d}{d \tau}
\left(
\begin{array}{c}
\delta x  \\
\delta y
\end{array}
\right)
\Bigg \vert _{(x,y)=(x_0,y_0)}
=
\left(
\begin{array}{cc}
0 & \frac{1}{2} \sqrt{\frac{\lambda}{V_0}} \\
\frac{\partial (dy / d \tau)}{\partial x} & \frac{\partial (dy / d \tau)}{\partial y}
\end{array}
\right)
\Bigg \vert _{(x,y)=(x_0,y_0)}
\left(
\begin{array}{c}
\delta x  \\
\delta y
\end{array}
\right) ,
\label{PA}
\end{equation}
where $(x_0,y_0)$ represents one of the points $(0,0)$, $(\pm 1,0)$ or $(\pm\infty,0)$.
A character of stationary points $(x_0,y_0)$ and a behavior of phase trajectories in their vicinity are determined by the eigenvalues of the matrix standing in the right-hand side of Eq.~(\ref{PA}). In particular, $(x_0,y_0)$ is stable if all real parts of eigenvalues are negative.

One can easily check that a $2\times 2$ matrix $\{(0,a),(b,c) \}$ has, generally speaking, two eigenvalues $(c \pm \sqrt{c^2 + 4ab})/2$. Both of them have negative real parts if and only if $c<0$ and $ab<0$. Moreover, since a configuration of phase trajectories is determined by  imaginary parts of eigenvalues, phase trajectories form spirals if $c^2+4ab<0$, and straight lines if $c^2+4ab \geq 0$. Using these algebraic facts, we can conclude that, in our case, the stationary points $(x_0,y_0)$ are stable if and only if
\begin{align}
\frac{\partial (dy/ d \tau)}{\partial x} \bigg \vert _{(x,y)=(x_0,y_0)} <0 \qquad
\mathrm{and} \qquad
\frac{\partial (dy/ d \tau)}{\partial y} \bigg \vert _{(x,y)=(x_0,y_0)} <0. \label{stability_cond}
\end{align}
In the table \ref{tab01} we enumerate all stationary points of the dynamical system (\ref{A1})-(\ref{A2}) and briefly characterize their stability. Below, let us discuss each stationary point in more detail.

\begin{table*}
	\caption{\label{tab01} Stationary points of the dynamical system (\ref{A1})-(\ref{A2}).}
	\begin{tabular}{ccccc}
		\hline\hline
		\textbf{No} &\quad & \textbf{~~~~~Stationary point~~~~~} &\quad & \textbf{~~~~~~~~~~Stability~~~~~~~~~~} \\
		\hline
		1. &\quad & $x=0$, $y=0$ &\quad & Unstable if $V_0\le1$ \\
		   &      &              &      & Stable if $V_0>1$ \\ \hline
		2. &\quad & $x=\pm1$, $y=0$ &\quad & Stable focuses \\ \hline
		3. &\quad & $x=\pm \infty$, $y=0$ &\quad &  Stable \\
		\hline\hline
	\end{tabular}
\end{table*}

\subsubsection{Stationary point $(x,y)=(0,0)$}

At the stationary point $(x,y)=(0,0)$ the elements of the matrix in Eq.~(\ref{PA}) are given as
\begin{align}
\frac{\partial (dy/ d \tau)}{\partial x} \bigg \vert _{(x,y)=(0,0)} &= 2 \frac{\sqrt{\lambda V_0}}{1-V_0 }, \label{stability_cond1}
\\
\frac{\partial (dy/ d \tau)}{\partial y} \bigg \vert _{(x,y)=(0,0)} &= -\sqrt{6 \pi \lambda \gamma}. \label{stability_cond2}
\end{align}
They satisfy the stability condition \Ref{stability_cond} provided $V_0 > 1$.
Therefore, the point $(0,0)$ is stable only if $V_0 > 1$, and unstable if $V_0<1$.

In case $V_0 = 1$ the expression \Ref{stability_cond1} for $dy/d \tau$ becomes indeterminate at $(x,y)=(0,0)$,
and hence we cannot even understand either this point is stationary or not.
Now, to characterize the point $(0,0)$,
it is necessary to consider its infinitesimal vicinity.
For example, at the points $(x,y)=(\pm \epsilon_1, \epsilon_2)$,
where $\epsilon_1$ and $\epsilon_2$ are small positive numbers of same order,
the derivatives $dx / d \tau$ and $\pm dy / d \tau$ are positive. At the same time,
at $(x,y)=(\epsilon_1 , \pm \epsilon_2)$, both $\pm dx / d \tau =0$ and $dy/d \tau$ are positive.
This means that the point $(x,y)=(0,0)$ is not stable (see Fig.~\ref{f2}).

\begin{figure}
\begin{minipage}[t]{0.5\columnwidth}
\begin{center}
\includegraphics[clip, width=0.97\columnwidth]{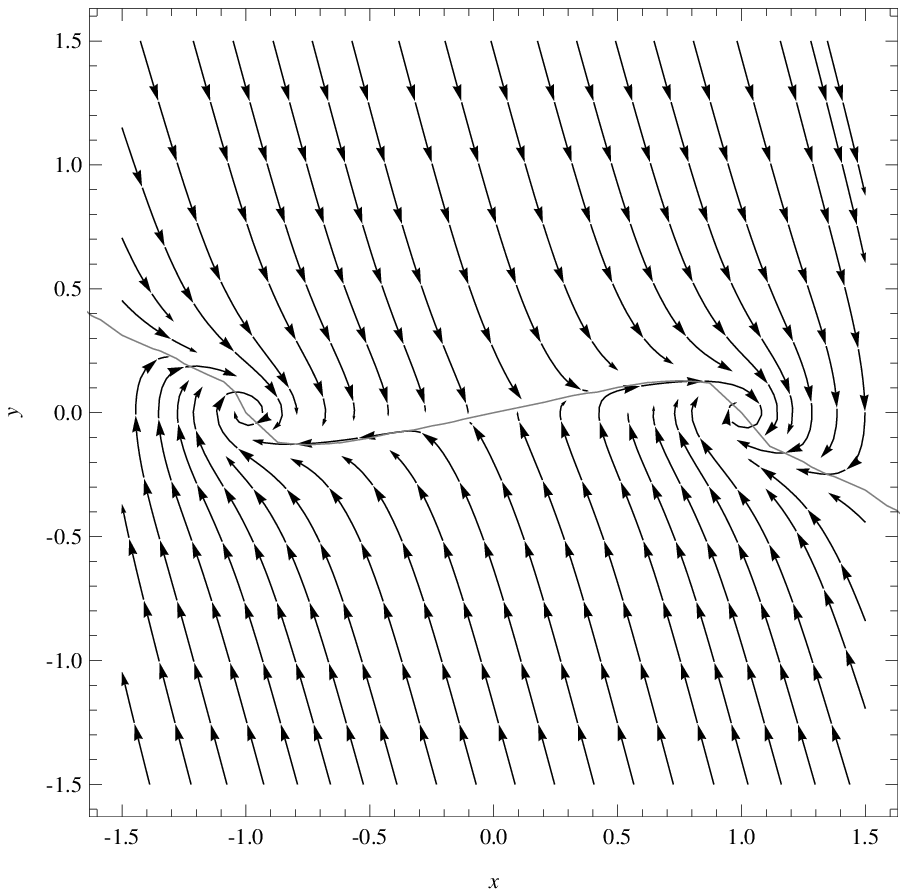}
\end{center}
\end{minipage}%
\begin{minipage}[t]{0.5\columnwidth}
\begin{center}
\includegraphics[clip, width=0.97\columnwidth]{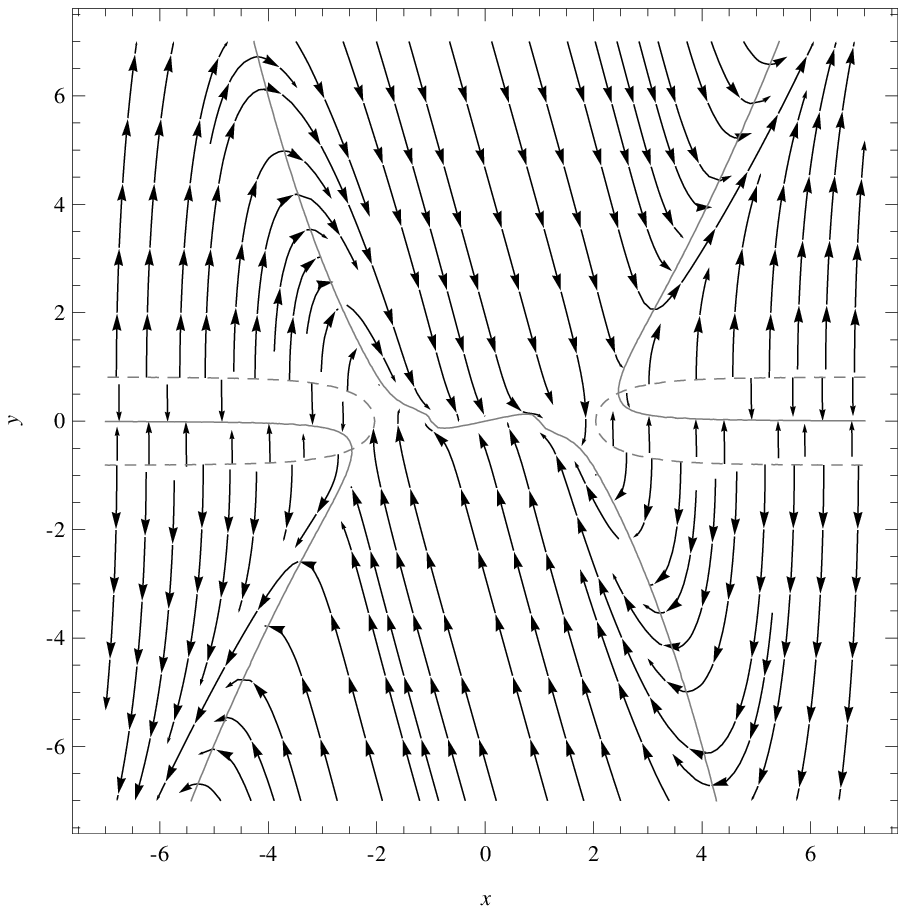}
\end{center}
\end{minipage}
\caption{Time evolution of the dynamical parameters $x= \phi / \phi _0$ and $y = \sqrt{8 \pi G \kappa} \dot \phi$
when $0<V_0<1$. The values of the constants $\gamma =0.5$, $\lambda = 0.2$, and $V_0 =0.1$ are assumed. The gray solid curves
and the gray dashed curves express $dy/ d \tau = 0$ and $dy/ d \tau = \pm \infty$, respectively. }
\label{f1}
\end{figure}

\subsubsection{Stationary points $(x,y)=(\pm 1,0)$}

Now, the elements of the matrix in Eq.~(\ref{PA}) are expressed as
\begin{align}
\frac{\partial (dy/ d \tau)}{\partial x} \bigg \vert _{(x,y)=(\pm 1,0)} &= -4 \sqrt{\lambda V_0 }, \label{fp10x} \\
\frac{\partial (dy/ d \tau)}{\partial y} \bigg \vert _{(x,y)=(\pm 1,0)} &= 0. \label{fp10y}
\end{align}
Since the right-hand side of Eq.~(\ref{fp10y}) equals to zero, and is not positive and not negative, we cannot determine either the points $(\pm 1, 0)$ are stable or not.
To characterize  stability by analytical methods, one can apply the centre manifold analysis (see e.g. \cite{Aulbach}). 
Otherwise, one can handle the problem numerically. Below we discuss an asymptotical behavior of phase trajectories near the stationary points. 
In particular, we consider the points $(\pm 1, 0)$ and show that they are stable attractive focuses. 

\begin{figure}
\begin{minipage}[t]{0.5\columnwidth}
\begin{center}
\includegraphics[clip, width=0.97\columnwidth]{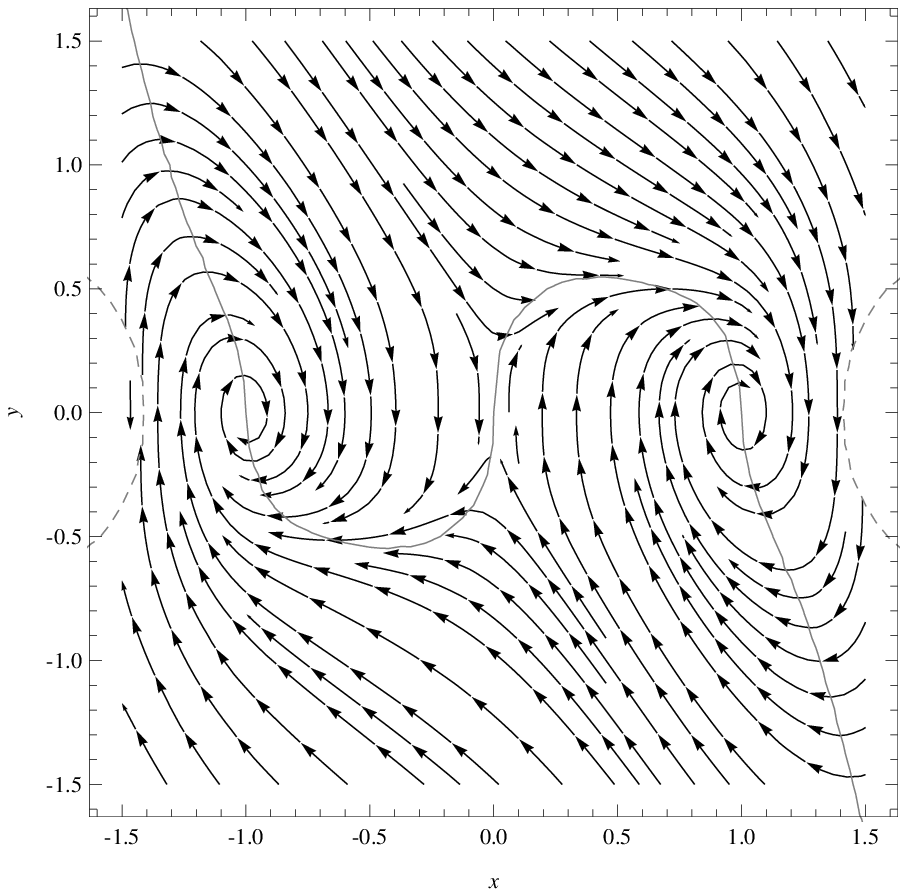}
\end{center}
\end{minipage}%
\begin{minipage}[t]{0.5\columnwidth}
\begin{center}
\includegraphics[clip, width=0.97\columnwidth]{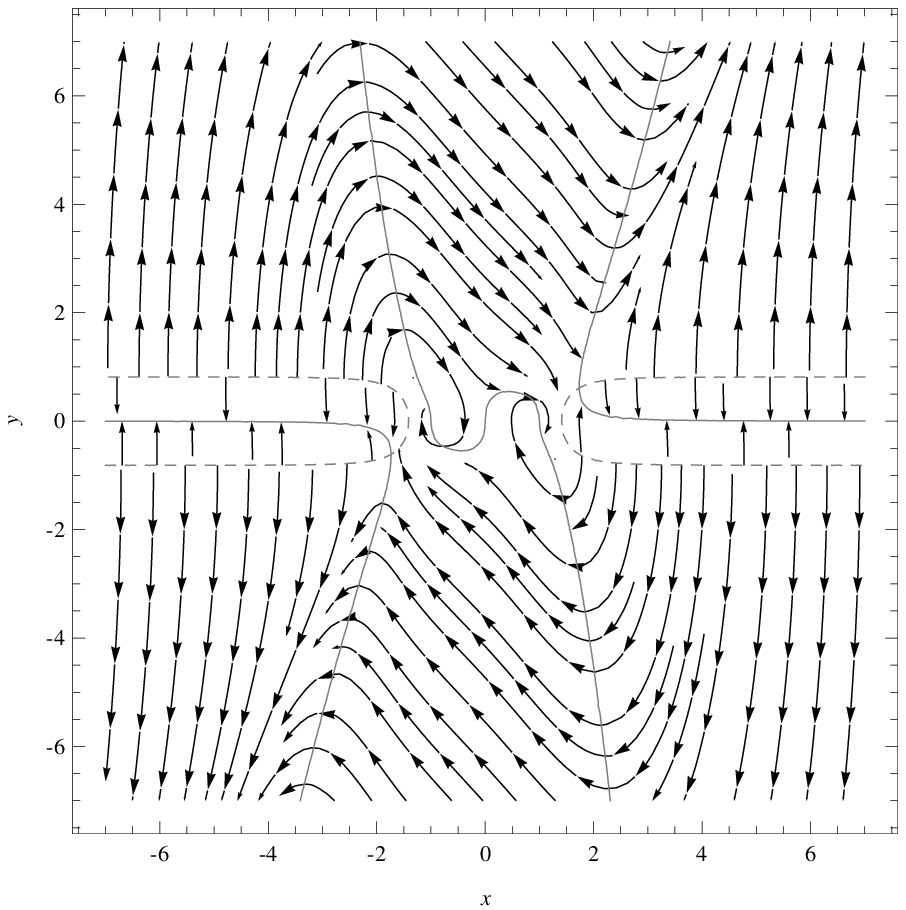}
\end{center}
\end{minipage}
\caption{Time evolution of the dynamical parameters $x= \phi / \phi _0$ and $y = \sqrt{8 \pi G \kappa} \dot \phi$
when $V_0=1$. The values of constants $\gamma =0.5$, $\lambda = 0.2$, and $V_0 =1$ are assumed.}
\label{f2}
\end{figure}
\begin{figure}
\begin{minipage}[t]{0.5\columnwidth}
\begin{center}
\includegraphics[clip, width=0.97\columnwidth]{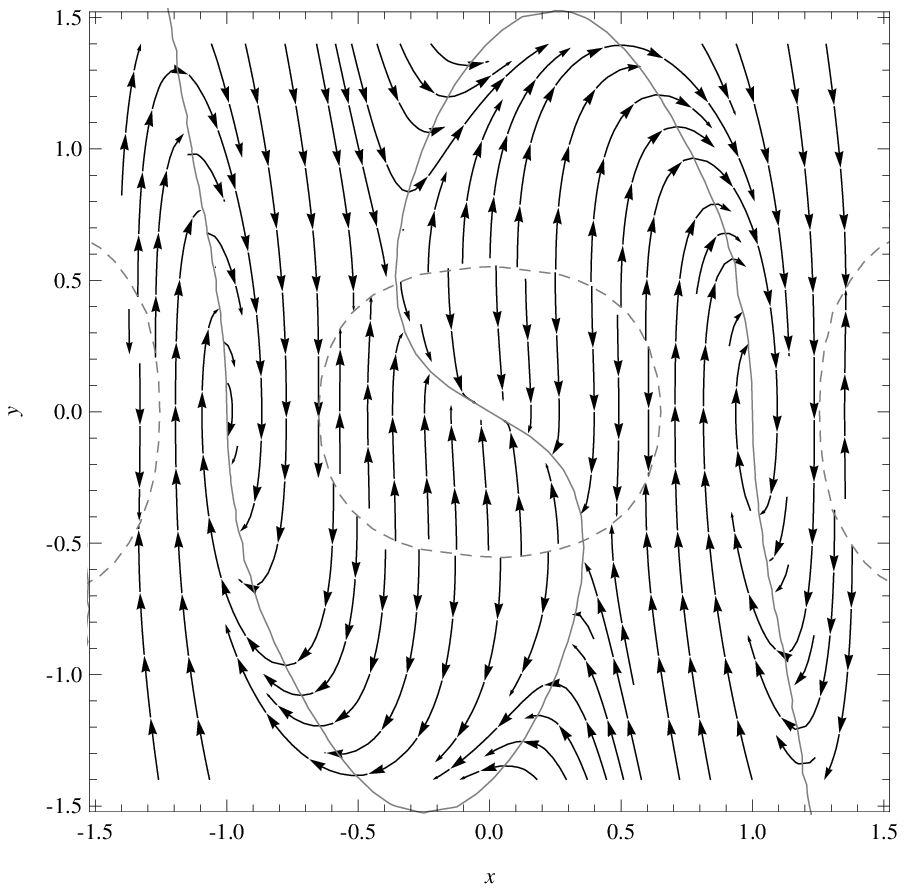}
\end{center}
\end{minipage}%
\begin{minipage}[t]{0.5\columnwidth}
\begin{center}
\includegraphics[clip, width=0.97\columnwidth]{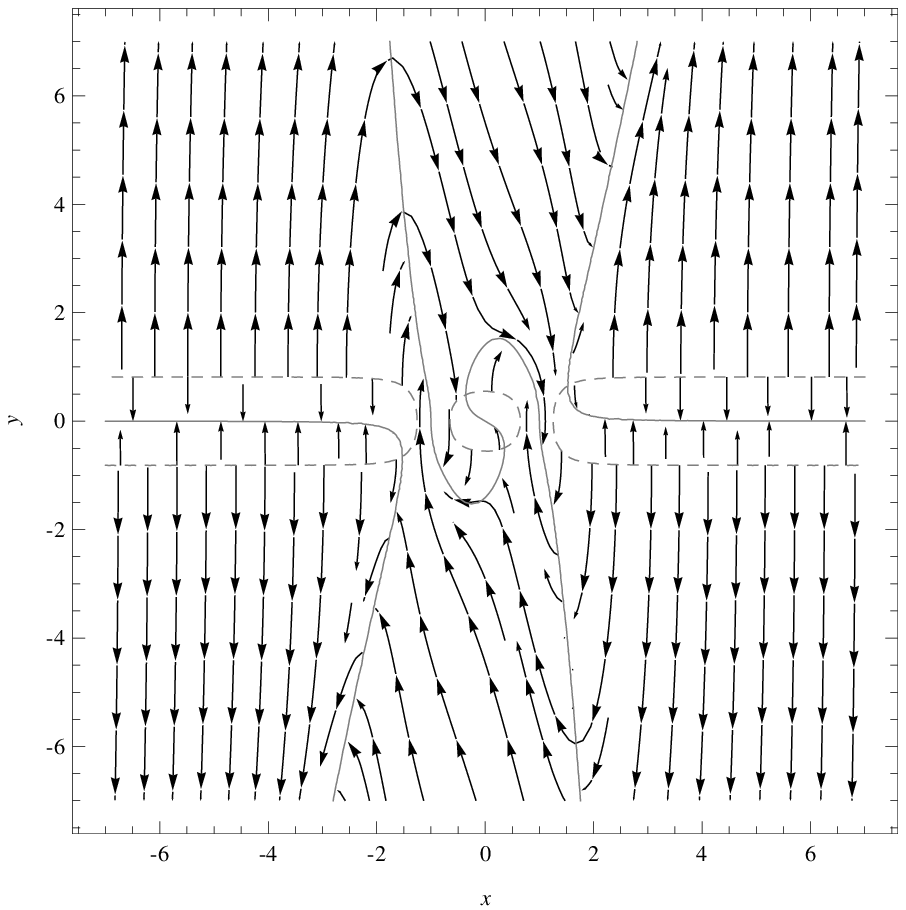}
\end{center}
\end{minipage}
\caption{Time evolution of the dynamical parameters $x= \phi / \phi _0$ and $y = \sqrt{8 \pi G \kappa} \dot \phi$
when $V_0>1$. The values of constants $\gamma =0.5$, $\lambda = 0.2$, and $V_0 =3$ are assumed.}
\label{f3}
\end{figure}

\subsubsection{Stationary points $(x,y)=(\pm\infty,0)$}

The elements of the matrix in Eq.~(\ref{PA}) at the points $(\pm\infty,0)$ can be found as
\begin{align}
\frac{\partial (dy/ d \tau)}{\partial x} \bigg \vert _{(x,y)=(\pm \infty,0)} &= - 2 \sqrt{\frac{\lambda}{V_0}}\lim_{L\to\pm\infty}\frac{1}{L^2}=-0, \label{fpL0x}\\
\frac{\partial (dy/ d \tau)}{\partial y} \bigg \vert _{(x,y)=(\pm \infty,0)} &=  - \sqrt{6 \pi \gamma \lambda}\lim_{L\to\pm\infty} L^2=-\infty. \label{fpL0y}
\end{align}
Both of the right-hand sides of Eqs.~(\ref{fpL0x}) and (\ref{fpL0y}) are negative.
Therefore, the stationary points $(\pm \infty,0)$ are stable.

\subsection{Critical points where $\frac{dy}{d \tau} =0$ and $\frac{dx}{d \tau} \not=0$ \label{dydtau0}}
In addition to stationary points, the dynamical system (\ref{A1}), (\ref{A2}) has other noteworthy ones. Among of them are critical points $(x,y)$, where one has $\frac{dy}{d \tau} =0$, while $\frac{dx}{d \tau} \not=0$.
In this case a tangent line to a phase trajectory, going through such the point, becomes to be horizontal, i.e. parallel to $x$-axis. Therefore, the critical point is a turning or inflection point of the phase trajectory. To find a critical point, we should equate the numerator in Eq. \Ref{A2} to zero:
\begin{align}
	 \sqrt{3 \pi \gamma \lambda V_0^{-1}}\, y \left [ 1+y^2-V_0 (x^2-1)^2 \right ] \sqrt{[y^2+2V_0(x^2-1)^2]\left({\textstyle\frac{3}{2}}y^2+1 \right)}
	 &\nonumber \\
	\textstyle
	+ 2 \sqrt{\lambda V_0} \left ( \frac{3}{2}y^2 +1 \right ) \left ( \frac{1}{2}y^2 +1 \right ) x(x^2-1)&=0.
	\label{crit}
\end{align}
This equation determines a curve on the phase diagram consisting of critical points, i.e. a critical curve.

In Figs.~\ref{f1}-\ref{f3} the critical curve is represented by solid lines. It is seen that, generally, this curve consists of three branches. Two of them are situated in the first and third quadrants, and the third one passes through the origin. One can check that the critical curve given by Eq. \Ref{crit} possesses the central symmetry
and has the following asymptotics:

In the limit $x\to\pm\infty$, $y\to \pm0$
\beq\label{critcurv1}
y \simeq \sqrt{\frac{2}{3\pi\gamma V_0}}\,\frac{1}{x^3}.
\eeq

In the limit $x\to\pm\infty$, $y\to\pm\infty$
\begin{equation}
y\simeq \pm(4\pi\gamma V_0)^{1/4}\, x^{3/2}.
\label{critcurv2}
\end{equation}

In the limit $x\to\mp\infty$, $y\to\pm\infty$
\begin{equation}
y\simeq -\frac{V_0}{\sqrt{2\pi \gamma}}\, x^3.
\label{critcurv3}
\end{equation}

\subsection{Critical points where $\frac{dy}{d \tau} =\pm \infty$ and $\frac{dx}{d \tau}$ is finite}
The other noteworthy points of the dynamical system (\ref{A1}), (\ref{A2}) are those where $\frac{dy}{d \tau} =\pm \infty$, while $\frac{dx}{d \tau}$ is finite.
It is necessary to emphasize that at such the points a tangent line to a phase trajectory takes a vertical position, i.e. parallel to $y$-axis, and the trajectory itself cannot be unambiguously extended over this point.

To find points where $\frac{dy}{d \tau} =\pm \infty$, we should equate the denominator in Eq. \Ref{A2} to zero, i.e. $\Delta=0$:
\beq
{1+\frac{3}{2}y^2+\frac{3}{2}y^4+V_0 (x^2-1)^2 \left ( \frac{3}{2}y^2 -1 \right )}=0.
\label{singular}
\eeq
This equation determines a curve on the phase diagram corresponding to the singularity $\frac{dy}{d \tau} =\pm \infty$. In Figs.~\ref{f1}-\ref{f3} these curves are shown as gray dashed lines.

Note that real-valued solutions of Eq.~(\ref{singular}) exist only if $V_0 \neq 0$.
Resolving Eq.~(\ref{singular}) with respect to $y$ gives
\begin{equation}
y^2 = - \frac{1}{2} \left [ 1 + V_0 (x^2-1)^2 \right ] + \frac{1}{2}
\sqrt{\left [ 1 + V_0 (x^2-1)^2 \right ]^2 + \frac{8}{3}[V_0 (x^2 -1)^2-1]}.
\label{singy}
\end{equation}
Values of $y$ are real provided $V_0 (x^2-1)^2 \geq 1$. In the case $V_0 (x^2-1)^2 = 1$ we have $x = \pm(1 \pm 1/ \sqrt{V_0})$. This relation gives two roots for $x$ if $0< V_0 <1$, three roots if $V_0 = 1$, and four roots if $V_0>1$.
Of course, there is a relation between the value of $V_0$ and the stability of the stationary point $(x,y)=(0,0)$.
We can see in Fig.~\ref{f3} that the point $(x,y)=(0,0)$ becomes stable
by surrounding an elliptic line of the singularities in $dy/d \tau$ (expressed as gray dashed lines).

Now, resolving Eq.~(\ref{singular}) with respect to $x$, we find
\begin{equation}
\label{singx}
(x^2-1)^2 = \frac{1+\frac{3}{2}y^2+\frac{3}{2}y^4}{V_0 \left ( 1- \frac{3}{2}y^2 \right )}.
\end{equation}
Solutions of this equation exist only in the range $-\sqrt{2/3}\leq y \leq \sqrt{2/3}$.
In particular, $y \simeq \pm \sqrt{2/3}$ is satisfied for large $x$.

\subsection{Asymptotics}

Significant features of a phase diagram are represented by its asymptotical structure, and so hereafter we discuss all possible asymptotics for phase trajectories of the dynamical system \Ref{A1}-\Ref{A2}.

\subsubsection{Trajectories approaching to the stationary point $(0,0)$}

As was shown, in case $V_0>1$ the stationary point $(0,0)$ is stable.
To obtain asymptotics for phase trajectories in the vicinity of $(0,0)$,
we substitute the expressions \Ref{stability_cond1} and \Ref{stability_cond2} into \Ref{PA} and after some algebra find
\beq\label{as1}
x(\tau)=e^{-\sqrt{\alpha}\tau}[C_1 e^{\sqrt{\beta}\tau}+C_2 e^{-\sqrt{\beta}\tau}],
\eeq
where
$\alpha={\frac32\pi\gamma\lambda}$,
$\beta={\frac32\pi\gamma\lambda-\frac{\lambda}{V_0-1}}$,
and $y(\tau)=\sqrt{4V_0/\lambda}\, \dot x(\tau)$.
It is explicitly seen that in case $V_0>1$ all phase trajectories tend to the point $(0,0)$, i.e. $x(\tau)\to 0$, $y(\tau)\to 0$ as $\tau\to\infty$. It is worth also noticing that the point $(0,0)$ is either an attractive focus if $\beta<0$, or an attractive node if $\beta>0$.

\subsubsection{Trajectories approaching to the stationary points $(\pm1,0)$}

Let us consider, without loss of generality, the stationary point $(1,0)$.
In its vicinity we have $|x-1| \ll 1$ and $|y| \ll 1$. Restoring the dimensional units and introducing an auxiliary scalar field $\psi\equiv \phi-\phi_0$, we can rewrite the latter conditions as $|\psi| \ll 1$ and $8\pi G\kappa\dot{\psi}^2 \ll 1$. The Higgs potential $V(\phi)=\frac{\lambda}{4}(\phi^2-\phi_0^2)^2$ becomes asymptotically to be quadratic, i.e. $V(\psi)= U_0\psi^2$, where $U_0=\lambda\phi_0^2$.
And now, after neglecting corresponding terms, the scalar field equation \Ref{phi2gen} takes the following approximate form:
\beq\label{aseqVgen}
\ddot\psi=-2\sqrt{3\pi G}\dot\psi \sqrt{\dot{\psi}^2+2V(\psi)}-V_\psi.
\eeq
It is worth emphasizing that this equation does not contain $\kappa$ and has the same form as in the theory of the usual minimally coupled scalar field. It has well-know asymptotics which are  represented as damped oscillations. In the particular case of the quadratic potential $V(\psi)=U_0\psi^2$ one has \cite{Star}
\beq\label{damposcilphi}
\psi_{t\to\infty}\approx \frac{\sin m t}{\sqrt{3\pi G}\,m t},
\eeq
where $m=\sqrt{2U_0}=\sqrt{2\lambda\phi_0^2}$ is a scalar mass.

\subsubsection{Trajectories approaching to the stationary points $(\pm\infty,0)$}

To find asymptotics in the vicinity of the stationary points $(\pm\infty,0)$,
we assume the following asymptotical behavior
$x\sim \tau^\alpha$ and $y\sim dx/d\tau\sim \tau^{\alpha-1}$
at $\tau\to\infty$, where $\alpha<1$.
Substituting these asymptotics into Eqs. \Ref{A1}, \Ref{A2}, we find
\beq\label{asphi3}
x(\tau)=\pm k\tau^{1/4}+O(\tau^{-3/4}), \quad
y(\tau)=\pm \sqrt{\frac{V_0}{4\lambda}}k\tau^{-3/4}+O(\tau^{-7/4}),
\eeq
where $k=(8\lambda/3\pi\gamma V_0^2)^{1/8}$. Now, eliminating $\tau$ gives
$y\approx \sqrt{\frac{2}{3\pi\gamma V_0}}\frac{1}{x^3}$. Note that the last expression represents  the asymptotic \Ref{critcurv1}, hence phase trajectories approximate to the critical curve $dy/d\tau=0$ at large times.

\subsubsection{Trajectories in the limit $x\to\pm\infty$ and $y\to\pm\infty$}

As one can see in Figs. 1-3, there exists a family of phase trajectories $(x(\tau),y(\tau))$ such that $x(\tau)\to\pm\infty$ and $y(\tau)\to\pm\infty$. Let us suppose that $x(\tau)$ takes the following asymptotical form: $x(\tau)\approx q(\tau_*-\tau)^{-\alpha}$ as $\tau\to\tau_*$, where $\tau_*$ is a finite moment of time, i.e. $\tau_*<\infty$. Substituting this asymptotics into \Ref{A1}, \Ref{A2}, we find
\beq\label{asphi4}
x(\tau)\simeq \pm\frac{q}{(\tau_*-\tau)^{2}}, \quad
y(\tau)\simeq \pm\frac{\sqrt{16q^2V_0/\lambda}}{(\tau_*-\tau)^{3}},
\eeq
where $q={14}(V_0/\pi\gamma\lambda^2)^{1/2}$. Then, eliminating $\tau$, we obtain the asymptotics $y\simeq \pm(16V_0/\lambda q)^{1/2}|x|^{3/2}$.

\subsection*{The case $\kappa < 0$}
\begin{figure}
\begin{minipage}[t]{0.5\columnwidth}
\begin{center}
\includegraphics[clip, width=0.97\columnwidth]{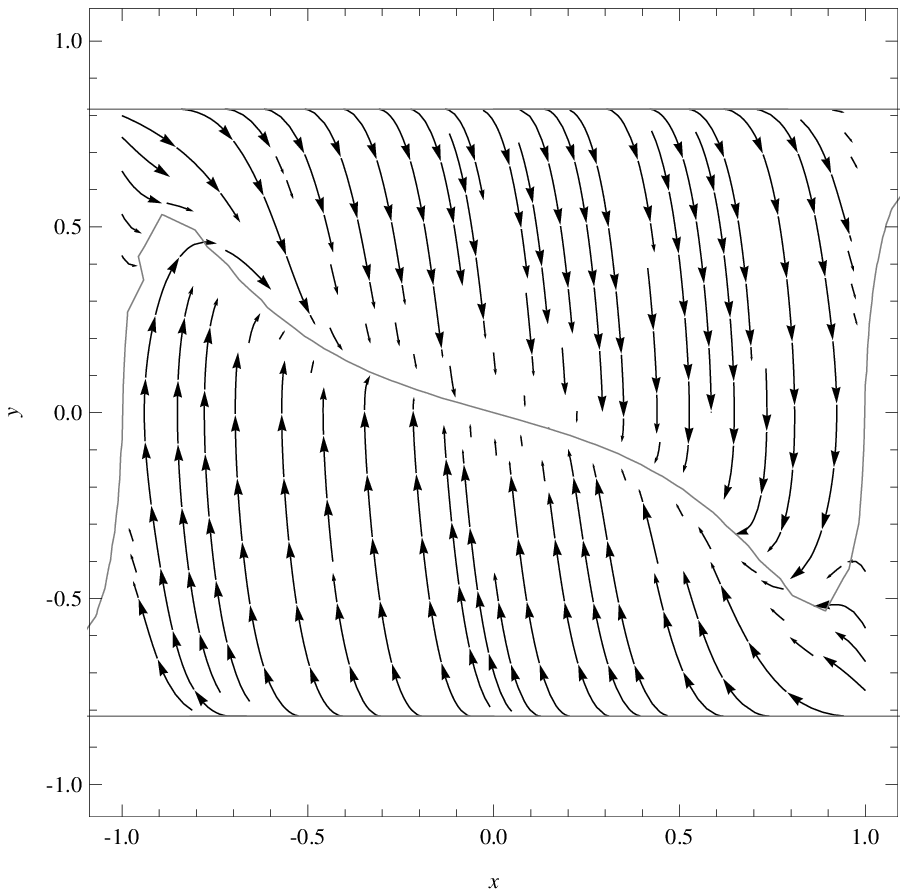}
\end{center}
\end{minipage}%
\begin{minipage}[t]{0.5\columnwidth}
\begin{center}
\includegraphics[clip, width=0.97\columnwidth]{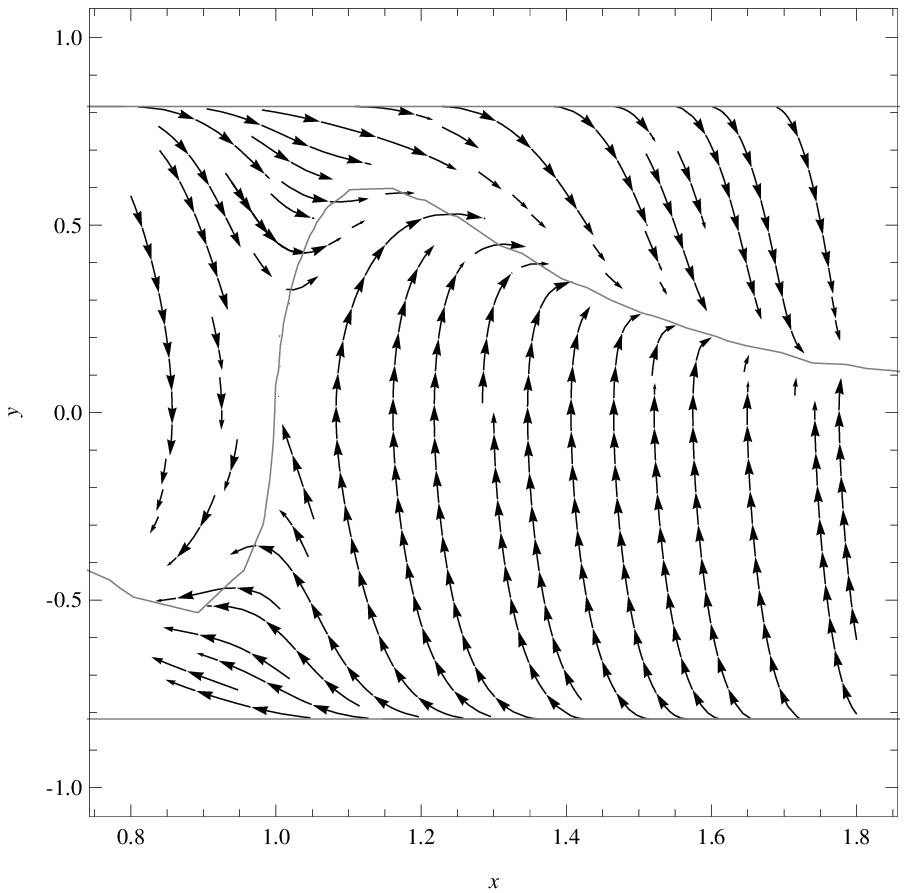}
\end{center}
\end{minipage}
\caption{Time evolution of the dynamical parameters $x = \phi / \phi _0$ and $y = \sqrt{8 \pi G |\kappa|} \dot \phi$ when $\kappa <0$. The values of constants $\gamma =0.5$, $\lambda = 0.2$, and $V_0 = 3$ are assumed.}
\label{f4}
\end{figure}

Assume that $\kappa$ is negative. Now, some of dimensionless values should be redefined as follows
\beq
y = \sqrt{8 \pi G |\kappa|}\, \dot \phi, \quad V_0=2\pi G|\kappa|\lambda\phi_0^4.
\eeq
Then, Eqs.~(\ref{A1}) and (\ref{A2}) are rewritten as
\begin{align}
\frac{dx}{d \tau} &= \frac{1}{2} \sqrt{\frac{\lambda}{V_0}}\, y,
\label{A1'} \\
\frac{dy}{d \tau} &= \frac{1}
{\Delta}
\left \{ - \sqrt{\frac{3 \pi \gamma \lambda}{V_0}} y \left [ 1-y^{2} +V_0 (x^2-1)^2 \right ]
\sqrt{[y^{2} +2V_0(x^2-1)^2]\left(1- \textstyle \frac{3}{2}y^{2} \right)}
\right. \nonumber \\
&\textstyle + 2 \sqrt{ \lambda V_0} \left (1- \frac{3}{2}y^{2}  \right ) \left ( 1- \frac{1}{2}y^{2} \right ) x(x^2-1) \Bigg \}.
\label{A2'}
\end{align}
where
\beq
\Delta=\textstyle 1-\frac{3}{2}y^{2}+\frac{3}{2}y^{4} +V_0 (x^2-1)^2 \left ( \frac{3}{2}y^{2} +1 \right ).
\eeq
From Eq. \Ref{A2'} it is seen that the dynamics of $y$ is restricted by the narrow range $-\sqrt{2/3} \leq y \leq \sqrt{2/3}$ in case $\lambda>0$.
The stationary points of the system \Ref{A1'}-\Ref{A2'} are the same as those for $\kappa >0$, i.e. $(0,0)$, $(\pm 1,0)$, and $(\pm \infty,0)$. However, their properties are different.
The elements of the matrix in Eq.~(\ref{PA}) at the stationary points are
\begin{align}
\frac{\partial (dy / d \tau)}{\partial x} \bigg \vert _{(x,y)=(0,0)} &= - 2 \frac{\sqrt{\lambda V_0} }{1+V_0}, \label{sc00a} \\
\frac{\partial (dy / d \tau)}{\partial y} \bigg \vert _{(x,y)=(0,0)} &= -\sqrt{6 \pi \gamma \lambda }, \label{sc00b} \\
\frac{\partial (dy / d \tau)}{\partial x} \bigg \vert _{(x,y)=(\pm 1,0)} &= 4 \sqrt{\lambda V_0 }, \label{sc10a} \\
\frac{\partial (dy / d \tau)}{\partial y} \bigg \vert _{(x,y)=(\pm 1,0)} &= 0, \label{sc10b} \\
\frac{\partial (dy / d \tau)}{\partial x} \bigg \vert _{(x,y)=(\pm \infty,0)} &=
-\sqrt{\frac{4\lambda}{V_0}}\lim_{L\to\pm\infty}\frac{1}{L^2}=-0, \label{sci0a} \\
\frac{\partial (dy / d \tau)}{\partial y} \bigg \vert _{(x,y)=(\pm \infty,0)} &= - \sqrt{6 \pi \gamma \lambda} \lim_{L\to\pm\infty}L^2=-\infty. \label{sci0b}
\end{align}
Comparing with the stability conditions \Ref{stability_cond}, we can conclude that the stationary points $(0,0)$ and $(\pm \infty, 0)$ are always stable, while the points $(\pm 1,0)$ are always unstable as long as $\lambda$ and $\gamma$ are positive. Here, it is worth noticing an ``unexpectedness'' of this result. Really, the points $(\pm1,0)$ correspond to the minimum of the Higgs potential $V(\phi)=\lambda/4(\phi^2-\phi_0^2)^2$, and so one can intuitively think that they should be stable. Vice versa, the point $(0,0)$ corresponds to the local maximum, and $(\pm \infty, 0)$ are infinitely large ``wings'' of the potential, hence they should be unstable. However, as we can see in Fig.~\ref{f4}, in the case of the negative kinetic coupling $\kappa<0$ the situation is drastically changed.

Let us discuss asymptotical properties of the dynamical system \Ref{A1'}-\Ref{A2'} in the case $\kappa<0$.

\subsubsection*{Trajectories approaching to the stationary point $(0,0)$}

Substituting the expressions \Ref{sc00a} and \Ref{sc00b} into \Ref{PA} gives us
the concrete values of the eigenvalues of the matrix in \Ref{PA}, therefore, we obtain
\beq\label{as1m}
x(\tau)=e^{-\sqrt{\alpha}\tau}[C_1 e^{\sqrt{\beta}\tau}+C_2 e^{-\sqrt{\beta}\tau}],
\eeq
where
$\alpha={\frac32\pi\gamma\lambda}$,
$\beta={\frac32\pi\gamma\lambda-\frac{\lambda}{V_0+1}}$,
and $y(\tau)=\sqrt{4V_0/\lambda}\, \dot x(\tau)$.
It is explicitly seen that all phase trajectories tend to the point $(x,y)=(0,0)$ in the vicinity of it.
It is worth also noticing that the point $(0,0)$ is either an attractive focus if $\beta<0$, or an attractive node if $\beta>0$.

\subsubsection*{Trajectories approaching to the stationary points $(\pm\infty,0)$}

To find asymptotics in the vicinity of the stationary points $(\pm\infty,0)$,
we assume the following asymptotical behavior
$x\sim \tau^\alpha$ and $y\sim dx/d\tau\sim \tau^{\alpha-1}$
at $\tau\to\infty$, where $\alpha<1$.
Substituting these asymptotics into Eqs. \Ref{A1'}, \Ref{A2'}, we find
\beq\label{asphi3m}
x(\tau)=\pm k\tau^{1/4}+O(\tau^{-3/4}), \quad
y(\tau)=\pm \sqrt{\frac{V_0}{4\lambda}}k\tau^{-3/4}+O(\tau^{-7/4}),
\eeq
where $k=(8\lambda/3\pi\gamma V_0^2)^{1/8}$. Note that these asymptotics coincide with those found in case of $\kappa >0$.

\subsection*{The case $\kappa = 0$}
\begin{figure}
\begin{minipage}[t]{0.5\columnwidth}
\begin{center}
\includegraphics[clip, width=0.97\columnwidth]{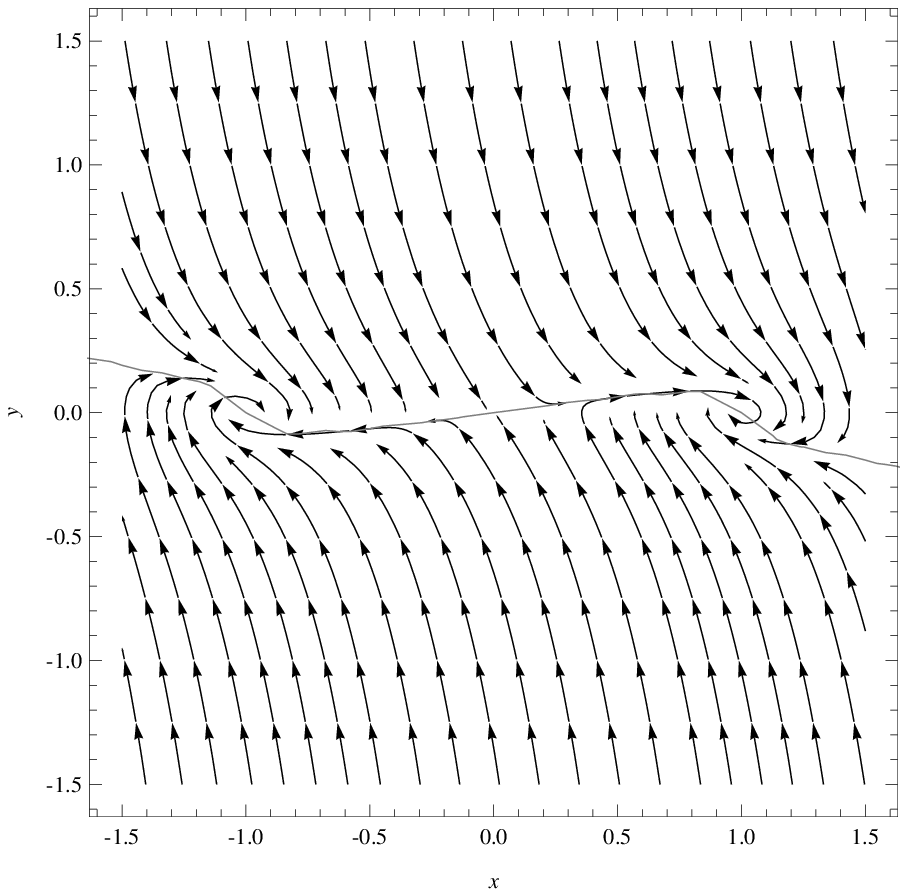}
\end{center}
\end{minipage}%
\begin{minipage}[t]{0.5\columnwidth}
\begin{center}
\includegraphics[clip, width=0.97\columnwidth]{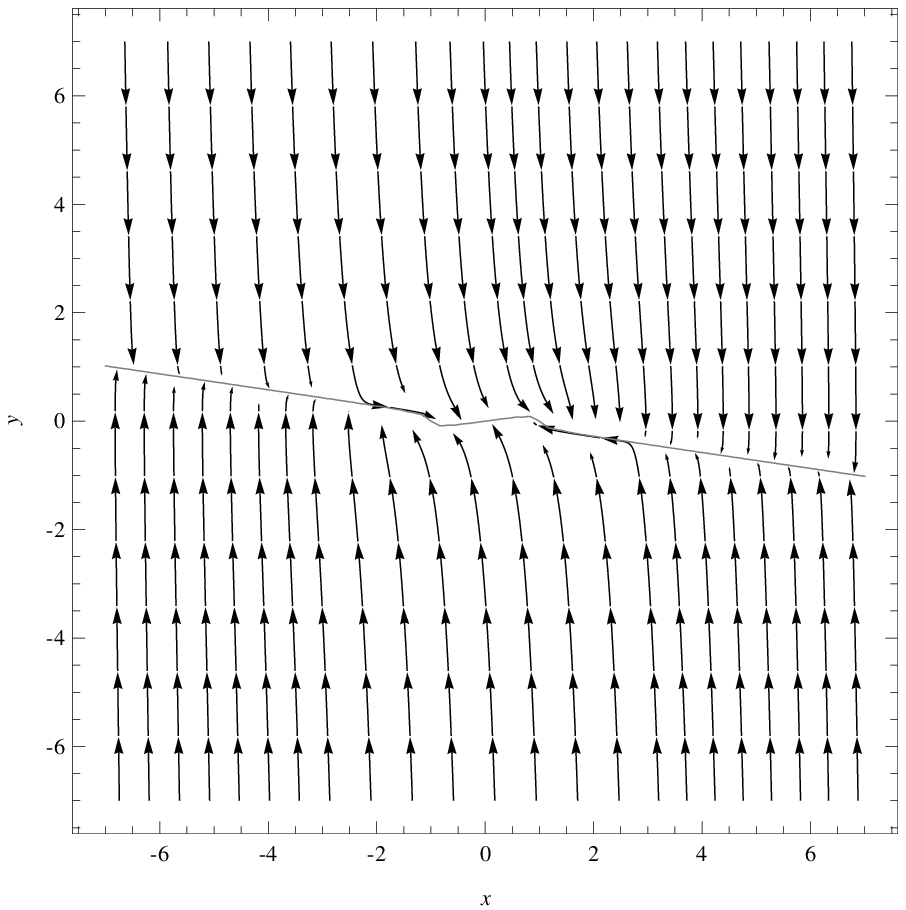}
\end{center}
\end{minipage}
\caption{Time evolution of the dynamical parameters $x = \phi / \phi _0$ and $y = \dot \phi / \phi _0 ^2$ in case $\kappa = 0$. The values of constants $\gamma =0.5$ and $\lambda = 0.2$ are assumed.}
\label{f5}
\end{figure}

To complete the consideration, we briefly review the case without kinetic coupling, i.e. $\kappa=0$.
Now, the field equation (\ref{phi2gen}) is simplified to
\begin{equation}
\ddot \phi = -2 \sqrt{3 \pi G} \dot \phi \sqrt{\dot \phi ^2 + 2V} - V_\phi .
\label{fe0}
\end{equation}
Introducing the dimensionless variable $x=\phi/\phi_0$ and $y = \dot \phi / \phi _0 ^2$,
we can rewrite Eq.~(\ref{fe0}) as follows
\begin{align}
\frac{dx}{d \tau} =& y, \label{A1''}\\
\frac{dy}{d \tau} =& -2 \sqrt{3 \pi \gamma} y \sqrt{y^{2} + \lambda (x^2 -1)^2 /2} - \lambda x (x^2-1). \label{A2''}
\end{align}
A phase diagram of the system (\ref{A1''})-(\ref{A2''}) is very simple (see Fig. \ref{f5}).
The only stationary points are $(x,y)=(0,0), (\pm 1,0)$.
One can easily check that $(0,0)$ is a saddle point, and $(\pm1,0)$ are stable focuses.

%
\section{Cosmological scenarios}
\label{CosS}
%
An evolution of the Hubble parameter $H$ is governed by the scalar field $\phi$ by means of the relation \Ref{constralphagen}. In the previous section we have described all possible asymptotical properties of the scalar field. Now, we will discuss corresponding asymptotics for the Hubble parameter.

\subsection*{\bf The case $\kappa>0$}
\subsection{Late-time de Sitter expansion} 
\begin{figure}
\begin{minipage}[t]{0.5\columnwidth}
\begin{center}
\includegraphics[clip, width=0.97\columnwidth]{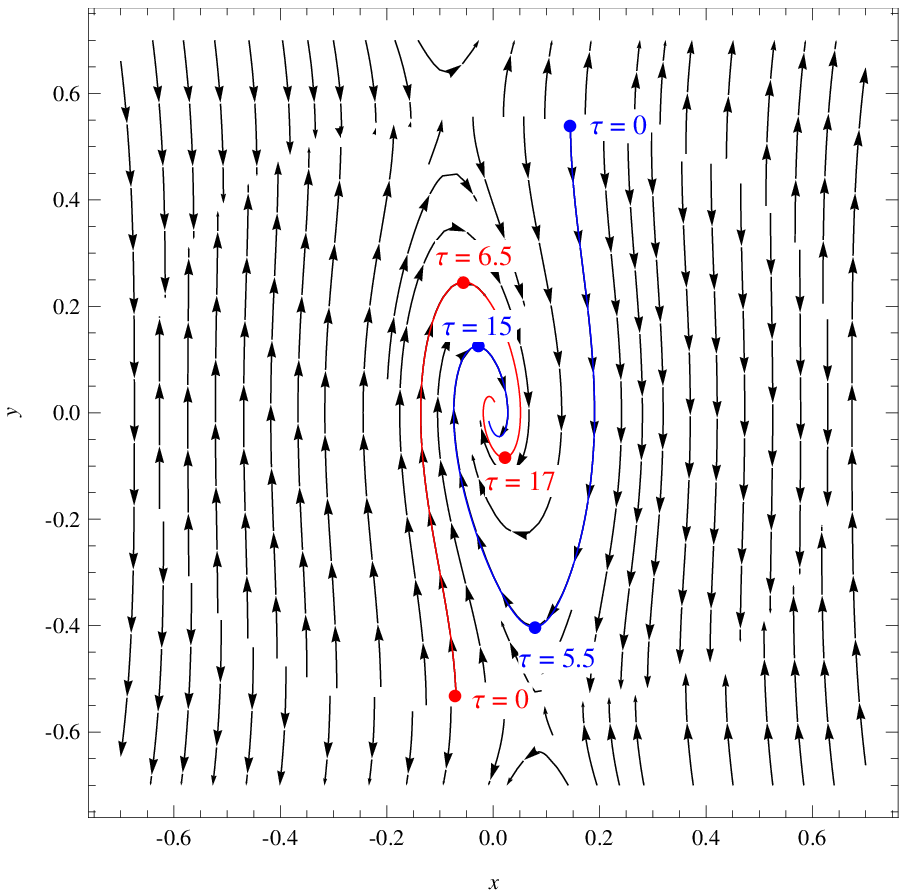}
\end{center}
\end{minipage}%
\begin{minipage}[t]{0.5\columnwidth}
\begin{center}
\includegraphics[clip, width=0.97\columnwidth]{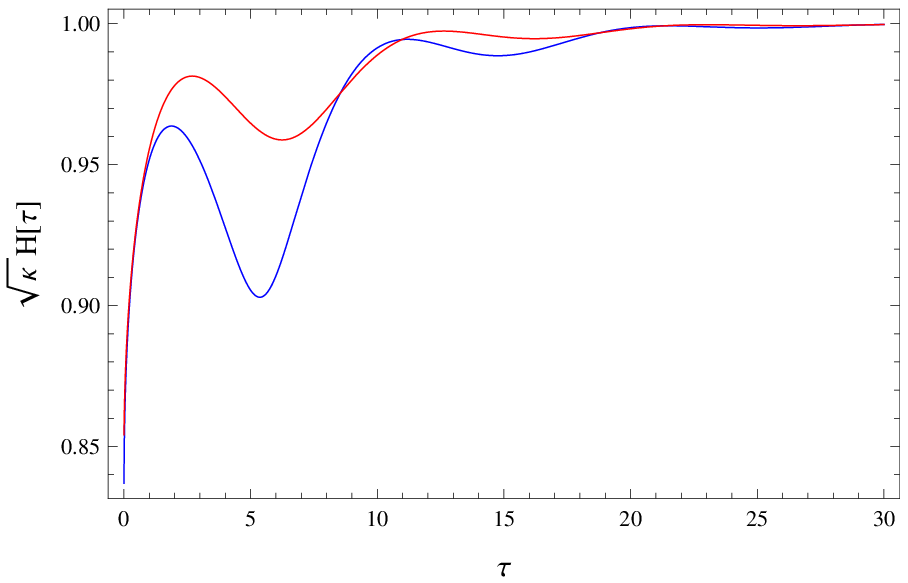}
\end{center}
\end{minipage}
\caption{Examples of the trajectories which go to the stationary point $(x,y)=(0,0)$ (left panel), and the corresponding time evolution of the Hubble parameter $H(t)$ (right panel). 
The values of constants are taken as $V_0=3$, $\gamma = 0.01$ and $\lambda =0.2$.}
\label{inf1}
\end{figure}

In case $V_0>1$ the stationary point $(0,0)$ of the dynamical system (\ref{A1}), (\ref{A2}) is stable and represents either an attractive node or focus depending on values of other parameters. The asymptotics for phase trajectories in the vicinity of $(0,0)$ are given by Eq. \Ref{as1}.
In the dimensional units the condition $V_0>1$ reads
\beq\label{condforkappa}
2 \pi G\kappa\lambda \phi _0 ^4 >1,
\eeq
and the scalar field has the following asymptotics:
\beq\label{asphi1}
\phi(t)=\phi_0 e^{-\sqrt{\tilde\alpha} t}[C_1 e^{\sqrt{\tilde\beta} t}+C_2 e^{-\sqrt{\tilde\beta} t}],
\eeq
where
$$
\tilde\alpha=\frac32\pi G\lambda\phi_0^4, \quad
\tilde\beta=\frac32\pi G\lambda\phi_0^4-\frac{\lambda\phi_0^2}{2\pi G\kappa\lambda\phi_0^4-1}.
$$
It is seen that in case $2 \pi G \kappa\lambda \phi _0 ^4 >1$ the scalar field and its derivatives go to zero at large times,
i.e. $\phi(t)\to 0$, $\dot\phi(t)\to 0$ as $t\to\infty$.
The corresponding value of the Higgs potential \Ref{Higgspotential} tends to its local maximum,
i.e. $V(0)=\lambda\phi_0^4/4$, and the Hubble parameter given by Eq. \Ref{constralphagen} tends to the constant value (see Fig.~\ref{inf1}):
\beq
H_{\infty}=\sqrt{\frac23\pi G\lambda\phi_0^4}.
\label{latetimeinf}
\eeq

Thus, we can conclude that the theory of gravity \Ref{action}
with the nonminimal kinetic coupling $\kappa>0$ and the Higgs potential could in principle provide a cosmological scenario with the late-time de Sitter epoch,
i.e. accelerated expansion of the Universe.
The parameters of the model should obey the condition $2 \pi G\kappa\lambda \phi _0 ^4 >1$.
 The cosmological constant in this epoch is $\Lambda_\infty=3H_\infty^2=2\pi G\lambda\phi_0^4$,
 and the Higgs potential reaches its local maximum $V(0)=\lambda\phi_0^4/4$.
 It is worth noticing that, finally, the value of $\Lambda_\infty$ is only determined by the parameters of the Higgs potential
 $\lambda$ and $\phi_0$ and does not depend on the coupling parameter $\kappa$.

\subsection{Oscillatory asymptotic}
\begin{figure}
\begin{minipage}[t]{0.5\columnwidth}
\begin{center}
\includegraphics[clip, width=0.97\columnwidth]{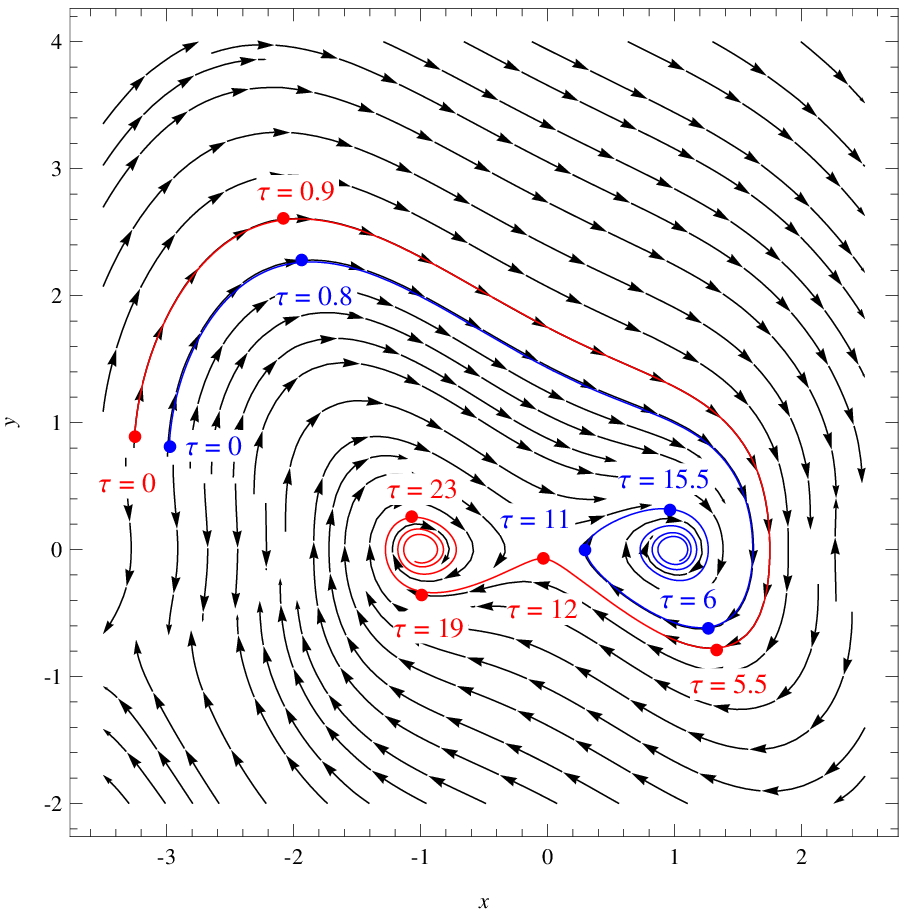}
\end{center}
\end{minipage}%
\begin{minipage}[t]{0.5\columnwidth}
\begin{center}
\includegraphics[clip, width=0.97\columnwidth]{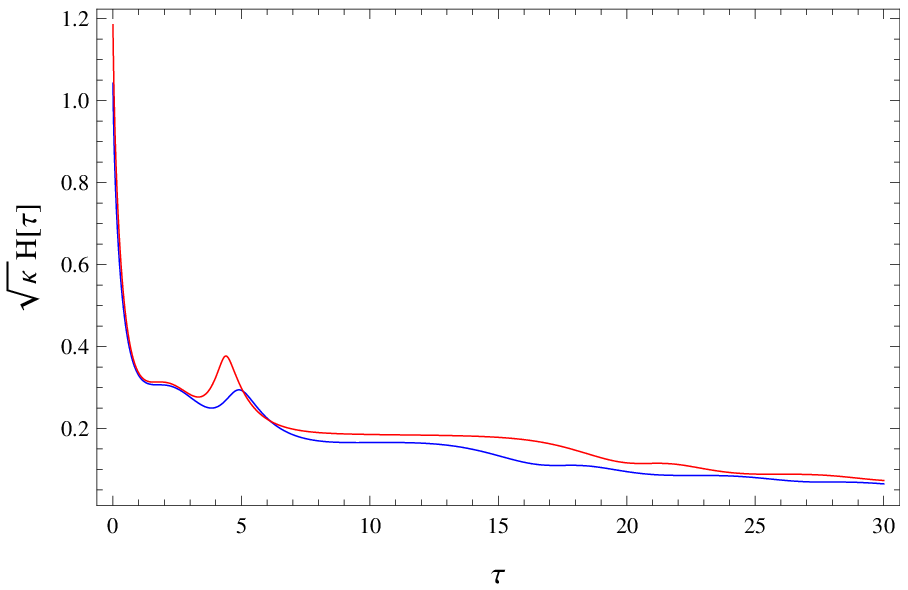}
\end{center}
\end{minipage}
\caption{Examples of the trajectories which go to the stationary point $(x,y)=(\pm 1,0)$ (left panel), and the corresponding time evolution of the Hubble parameter $H(t)$ (right panel). 
The values of constants are taken as $V_0=0.1$, $\gamma = 0.01$ and $\lambda =0.2$.}
\label{osci}
\end{figure}

An asymptotical behavior of the scalar field in the vicinity of stable attractive focuses $(\pm1,0)$ is represented as damped oscillations \Ref{damposcilphi}.
Now, from Eq. \Ref{constralphagen} we have the following asymptotics for the Hubble parameter:
\beq\label{damposcilH}
H_{t\to\infty}\approx H_{MD}(t)\,\left[1-\frac{\sin 2mt}{2mt}\right],
\eeq
where $m=\sqrt{2\lambda\phi_0^2}$ is a scalar mass and $H_{MD}(t)=2/3t$ is
the Hubble parameter in the matter-dominated Universe filled with nonrelativistic matter with $p\ll\rho$.
The oscillating behavior with damping can be seen in Fig.~\ref{osci}.

\subsection{Big Rip scenario}
\begin{figure}
\begin{minipage}[t]{0.5\columnwidth}
\begin{center}
\includegraphics[clip, width=0.97\columnwidth]{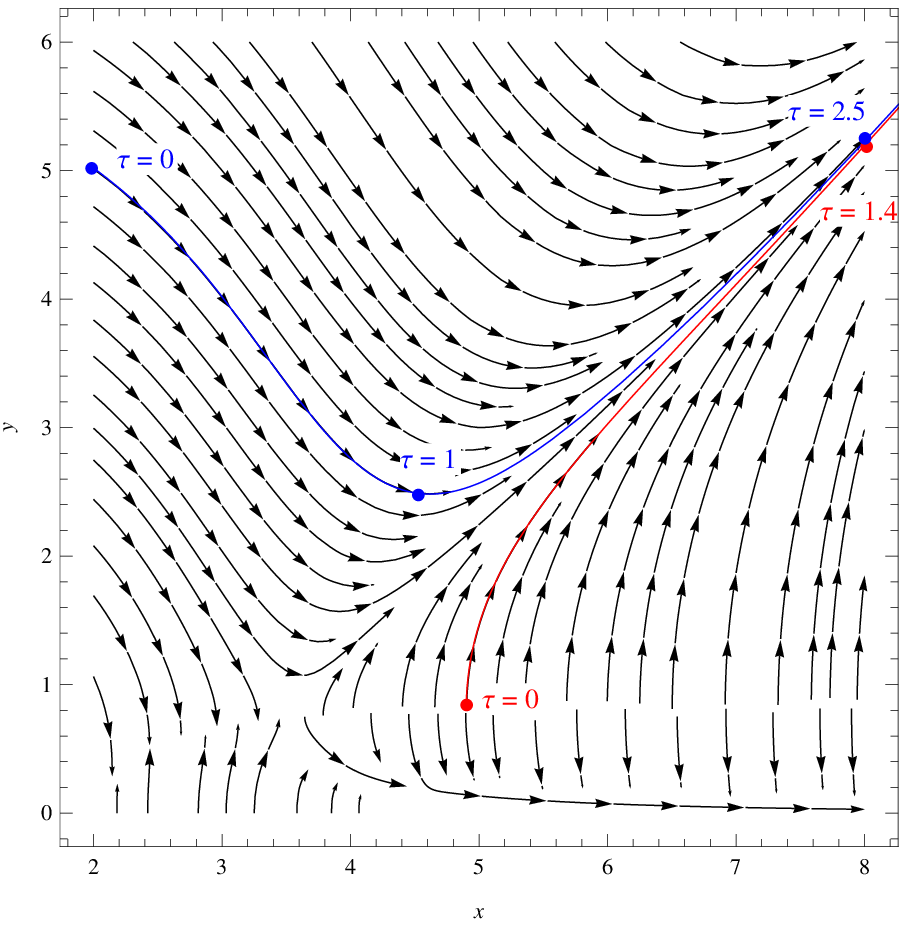}
\end{center}
\end{minipage}%
\begin{minipage}[t]{0.5\columnwidth}
\begin{center}
\includegraphics[clip, width=0.97\columnwidth]{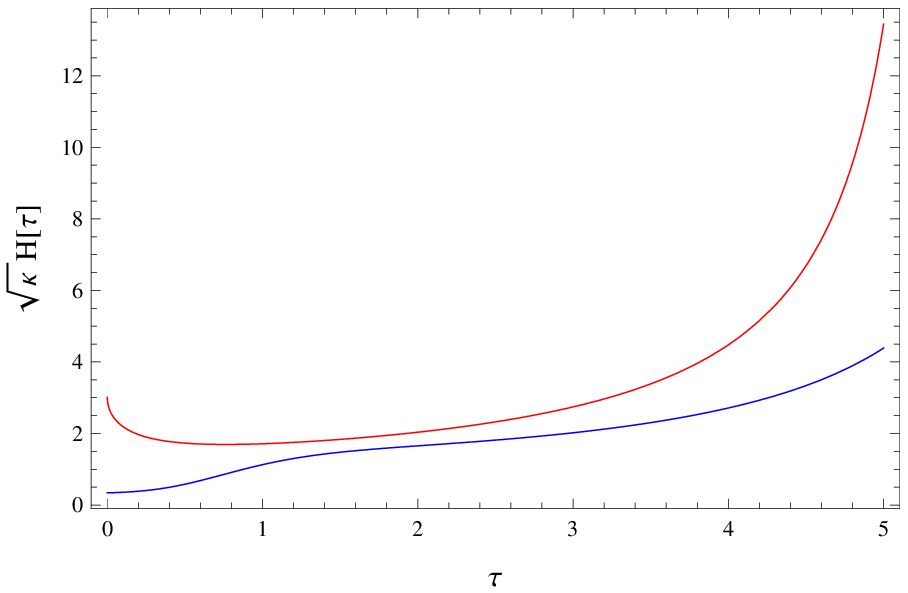}
\end{center}
\end{minipage}
\caption{Examples of the trajectories which approach the asymptote $y = (4 \pi ^2 \gamma V_0)^{1/4}|x|^{3/2}$ (left panel), and the corresponding time evolution of the Hubble parameter $H(t)$ (right panel). 
The values of constants are taken as $V_0=0.1$, $\gamma = 0.01$ and $\lambda =0.2$.
The time of singularity $\tau _* = \phi _0 t_*$ is about $5.5$ for red line and is about $6.5$ for blue line. }
\label{BR}
\end{figure}

As was shown, on the phase diagram there are phase trajectories $(x(\tau),y(\tau))$
such that $x(\tau)\to\pm\infty$ and $y(\tau)\to\pm\infty$ as $\tau\to\tau_*$, where $\tau_*<\infty$
is a finite moment of time.
The corresponding asymptotical behavior of $x(\tau)$ and $y(\tau)$ is given by Eq. \Ref{asphi4}.
In the dimensional units we have the following asymptotical behavior of the scalar field in the limit $t\to t_*$:
\beq
\phi(t)\simeq \sqrt{\frac{392\kappa}{\lambda}}\frac{1}{(t_*-t)^2}.
\label{BigRip-phi}
\eeq
The asymptotics for the Hubble parameter can be found from Eq. \Ref{constralphagen} as follows
\beq
H^2(t)\simeq \frac{49}{9} \frac{1}{(t_*-t)^2}.
\label{BigRip-H}
\eeq
Correspondingly, the scale factor asymptotically behaves as
\beq
a(t) \propto \frac{1}{(t_* -t)^{7/3}}.
\label{BigRip-a}
\eeq
Note that both the scale factor $a(t)$ and the Hubble parameter $H(t)$
go to infinity as the time tends to some finite value, $t\to t_*$. This type of a cosmological singularity is known as the Big Rip scenario.
Trajectories which can realize the Big Rip scenario are depicted in Fig.~\ref{BR}.

Note also that in the limit $\phi\to\infty$ the Higgs potential takes the power-law form $V(\phi)\approx \frac{\lambda}{4}\phi^4$. Cosmological dynamics with the nonminimal kinetic coupling and a power-law potential was studied, and the Big Rip scenario was represented by Eq. (28) in Ref. \cite{SkuSusTop:2013}.

\subsection{Little Rip scenario}
\begin{figure}
\begin{minipage}[t]{0.5\columnwidth}
\begin{center}
\includegraphics[clip, width=0.97\columnwidth]{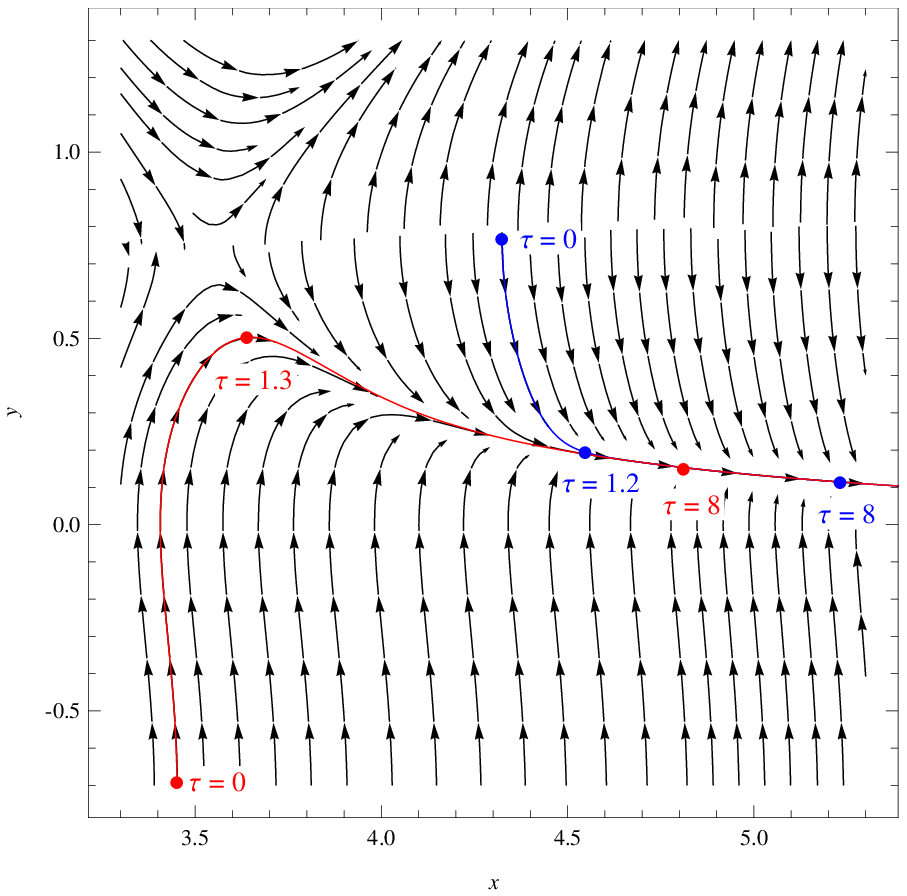}
\end{center}
\end{minipage}%
\begin{minipage}[t]{0.5\columnwidth}
\begin{center}
\includegraphics[clip, width=0.97\columnwidth]{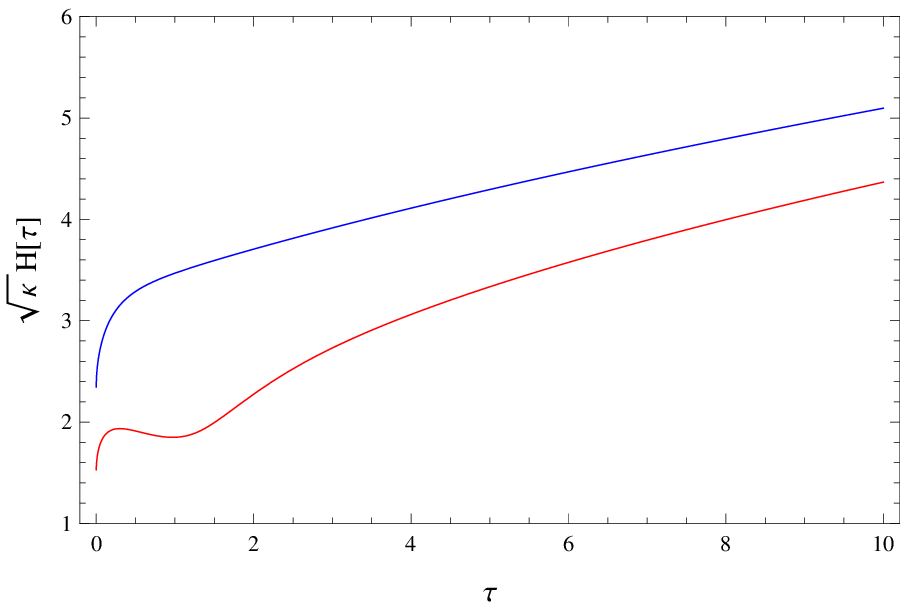}
\end{center}
\end{minipage}
\caption{Examples of the trajectories which approach the asymptote $y = [2/ (3 \pi \gamma V_0)]^{1/2}x^{-3}$ (left panel), and the corresponding time evolution of the Hubble parameter $H(t)$ (right panel). 
The values of constants are taken as $V_0=0.1$, $\gamma = 0.01$ and $\lambda =0.2$.}
\label{largex}
\end{figure}

In the vicinity of the stationary points $(\pm\infty,0)$ the asymptotics for $x(\tau)$ and $y(\tau)$ are given by Eq. \Ref{asphi3}.
Respectively, in the dimensional units we have the following asymptotical behavior of the scalar field in the limit $t\to \infty$:
\beq
\phi(t)\simeq (2/3\pi^3 G^3\lambda\kappa^2)^{1/8}\, t^{1/4},
\label{LittleRip-phi}
\eeq
that is asymptotically $\phi(t)\propto t^{1/4}$.
The corresponding asymptotic for the Hubble parameter can be found from Eq. \Ref{constralphagen} as follows
\beq
H(t)\simeq (8\lambda/27\pi G\kappa^2)^{1/4}\, t^{1/2},
\label{LittleRip-H}
\eeq
so that $H(t)\propto t^{1/2}$. Then, the scale factor asymptotically behaves as
\beq
a(t) \propto e^{k t^{3/2}},
\label{LittleRip-a}
\eeq
with $k=(2/3)(8\lambda/27\pi G\kappa^2)^{1/4}$. Such the behavior of the scale factor
means that the expansion of the Universe is running even faster than the inflationary (i.e. quasi-de Sitter) expansion. Note that the cosmological scenario where the Hubble parameter tends to infinity in the remote future, i.e. $H(t)\to\infty$ as $t\to\infty$, is known as the Little Rip \cite{LittleRip}. The trajectories which can realize the Little Rip expansion of the Universe are shown in Fig.~\ref{largex}.

\subsection{Is the slow-roll inflation possible?}
As is well known, the usual slow-roll conditions for the minimally coupled scalar field read
$\dot\phi^2/2\ll V(\phi)$ and $\ddot\phi \ll 3H\dot\phi$. Here, the first condition means that the kinetic energy is much less than the potential one, while the second condition says that the `viscosity' given by the term $3H\dot\phi$ is very high.
In the slow-roll approximation, the field equations are reduced to
$3H^2\simeq 8\pi G V(\phi)$ and $\dot\phi\simeq-\frac{V_\phi}{3H}$,
and, as the consequence, one additionally has
$-{\dot H}/{H^2}\ll 1$. The last condition provides an exponential (de Sitter) expansion of the Universe.

In the theory \Ref{action} with the nonminimal kinetic coupling, the field equations have the modified form \Ref{genfieldeq}, and hence the slow-roll conditions should be also modified. In particular, we can now change the condition $\dot\phi^2/2\ll V(\phi)$ as follows
\beq\label{modkinencond}
\textstyle\dot\phi^2|1-9\kappa H^2|\ll 2V(\phi),
\eeq
so that from Eq. \Ref{00cmpt} we obtain
\beq\label{slowrollrel1}
3H^2\simeq 8\pi G V(\phi).
\eeq

Note that in the paper \cite{Germani:2010gm} it was suggested to use the extra condition
\beq
\label{extracond}
H^2 \gg \frac{1}{9|\kappa|}
\eeq
for the model with the quartic potential. Here, following Ref. \cite{Germani:2010gm}, we will also request to obey the inequality \Ref{extracond}. Then, the condition \Ref{modkinencond} reduces to
\beq\label{modkinencond2}
\textstyle 9|\kappa| H^2 \dot\phi^2\ll 2V(\phi).
\eeq
Using the relation \Ref{slowrollrel1}, we can rewrite the conditions \Ref{extracond} and \Ref{modkinencond2} in the following form:
\beq\label{extracond2}
V(\phi) \gg \frac{1}{24\pi G |\kappa|},
\eeq
and
\beq\label{modkinencond3}
\frac12\dot\phi^2\ll \frac{1}{24\pi G |\kappa|}. \eeq
It is worth noticing that these conditions do not contain explicitly $H(t)$.

By using the dimensionless values, the new slow-roll conditions \Ref{extracond2} and \Ref{modkinencond3} can be written as
\beq
3V_0(x^2-1)^2 \gg 1, \label{NHIcond1}
\eeq
and
\beq
\textstyle\frac32 y^2 \ll 1. \label{NHIcond2}
\eeq
These conditions could be only satisfied in the vicinity of the critical points $(\pm\infty, 0)$, where $|x|\gg 1$ and $|y|\ll 1$, and, as was shown above,
an expansion scenario of the Universe in this case is the Little Rip.
To be specific, the conditions \Ref{NHIcond1} and \Ref{NHIcond2} restrict values of $x$ and $y$ to the regions being situated deep inside the left and right branches of the `critical' curve $dy/d\tau = \pm \infty$ (see Figs. \ref{f1}-\ref{f3}). Therefore, the Little Rip is realized because all trajectories in that region are approaching to the asymptote $y = [2/ (3 \pi \gamma V_0)]^{1/2}x^{-3}$.

In addition to \Ref{extracond2} and \Ref{modkinencond3}, we assume that the slow-roll conditions
\beq\label{slowrollcond3}
|\ddot\phi|\ll 3H|\dot\phi|
\eeq
and
\beq\label{slowrollcond4}
\frac{|\dot H|}{H^2}\ll 1
\eeq
are also fulfilled. Then, from Eq. \Ref{eqmocosm} we find
\beq
\dot\phi \simeq \frac{V_\phi}{9\kappa H^3}.
\eeq

The Little Rip asymptotics are given by Eqs. \Ref{LittleRip-phi}, \Ref{LittleRip-H}, \Ref{LittleRip-a}. In particular,
$\phi\sim t^{1/4}$ and $H\sim t^{1/2}$. Now, one can check that the Little Rip asymptotics obey all modified slow-roll conditions obtained above.

Summarizing, we conclude that the modified slow-roll conditions for the scalar field with
the nonminimal kinetic coupling consist of the following inequalities: $12\pi G\kappa\dot\phi^2\ll 1$, $24\pi G\kappa V(\phi)\gg 1$,
$|\ddot\phi|\ll 3H|\dot\phi|$, and $|\dot H|/H^2\ll 1$. In contrast to the usual slow-roll conditions which provide an exponential (de Sitter)
expansion of the Universe, the modified conditions lead to the Little Rip scenario.
It worth noticing that this result does not support
the assumption given in Ref. \cite{Germani:2010gm} that the slow rolling quasi-de Sitter expansion is realized if the slow-roll conditions are satisfied.
However, as we can see, a more thorough investigation reveals a qualitatively different behavior of the Universe expansion
under the modified slow-roll conditions.

\subsection*{\bf The case $\kappa<0$}
\begin{figure}
\begin{minipage}[t]{0.5\columnwidth}
\begin{center}
\includegraphics[clip, width=0.97\columnwidth]{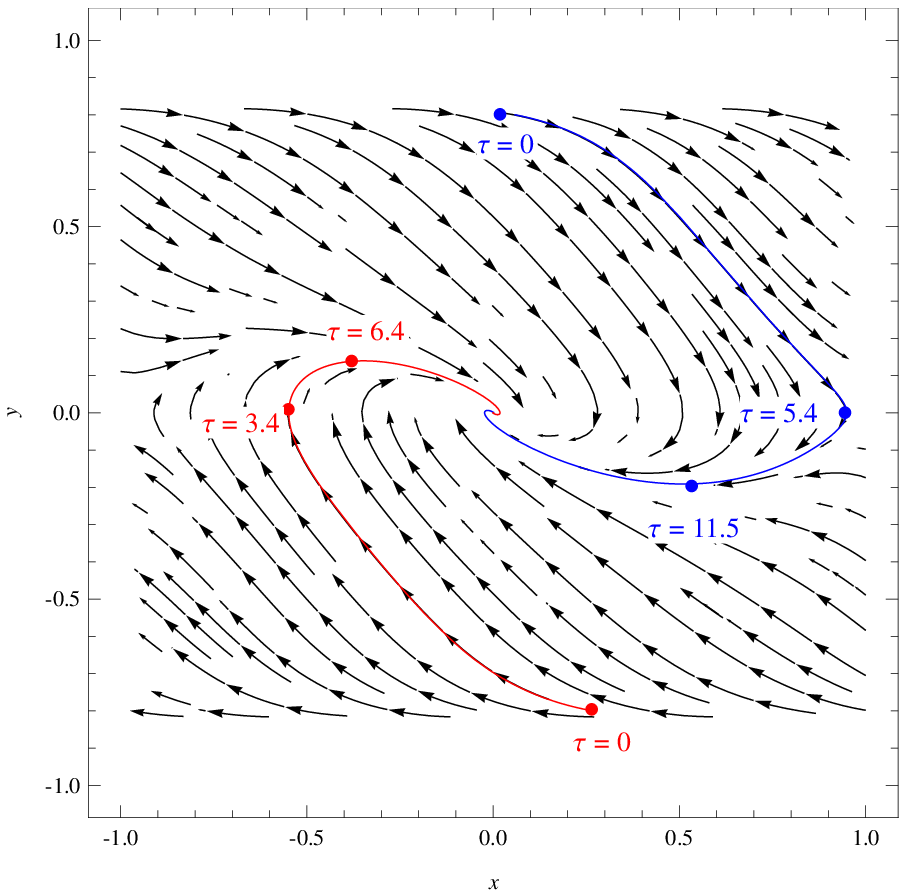}
\end{center}
\end{minipage}%
\begin{minipage}[t]{0.5\columnwidth}
\begin{center}
\includegraphics[clip, width=0.97\columnwidth]{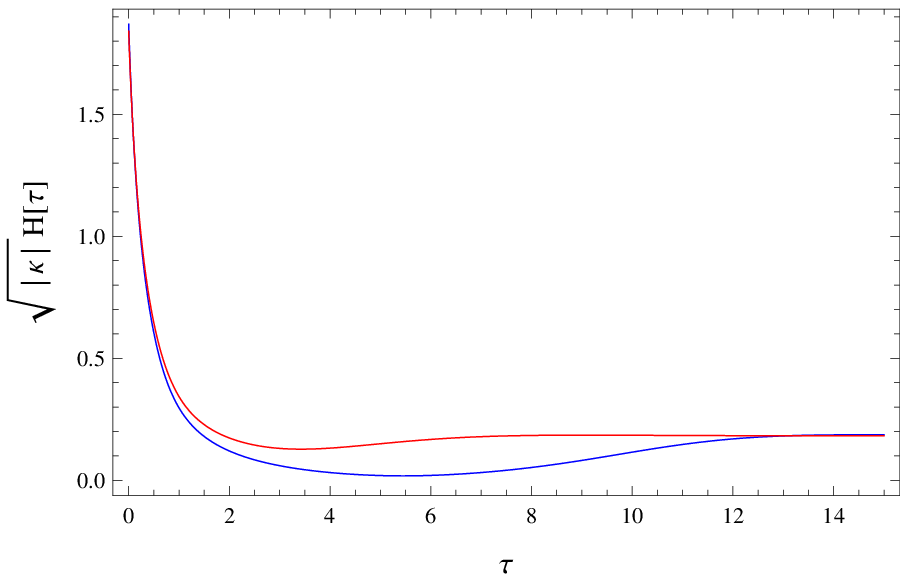}
\end{center}
\end{minipage}
\caption{Examples of the trajectories which go toward the point $(x,y)=(0,0)$ when $\kappa$ is negative (left panel), and the corresponding time evolution of the Hubble parameter $H(t)$ (right panel). 
The values of constants are taken as $V_0=0.1$, $\gamma = 0.1$ and $\lambda =0.2$.}
\label{nkt1}
\end{figure}

\begin{figure}
\begin{minipage}[t]{0.5\columnwidth}
\begin{center}
\includegraphics[clip, width=0.97\columnwidth]{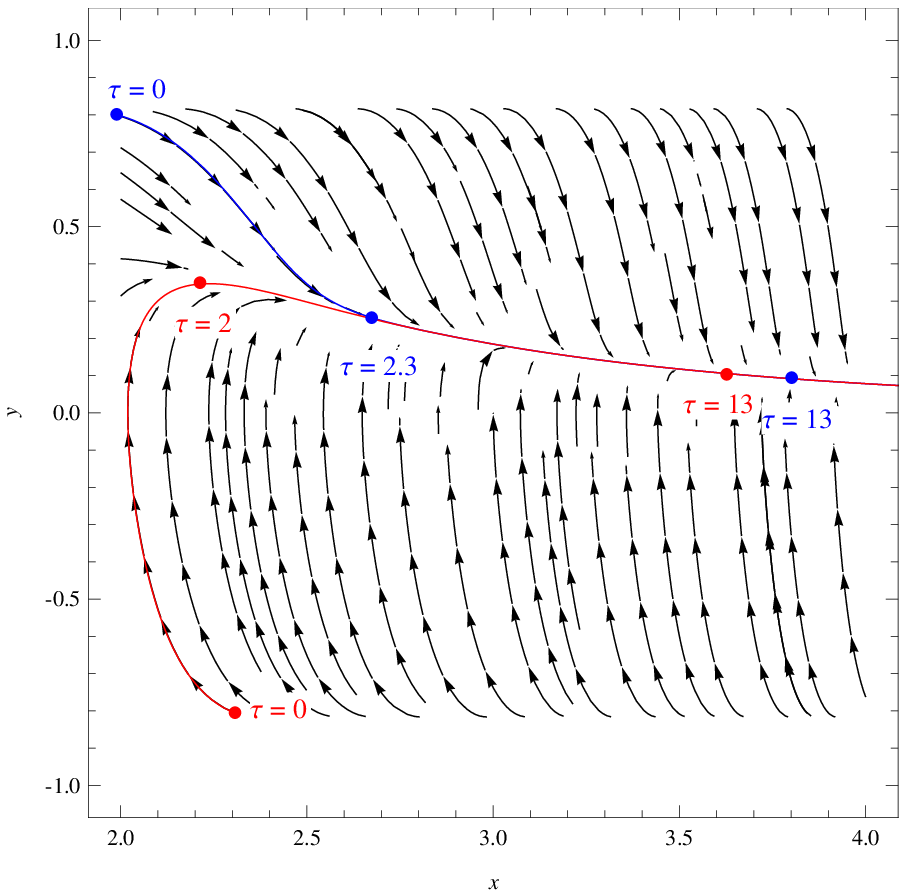}
\end{center}
\end{minipage}%
\begin{minipage}[t]{0.5\columnwidth}
\begin{center}
\includegraphics[clip, width=0.97\columnwidth]{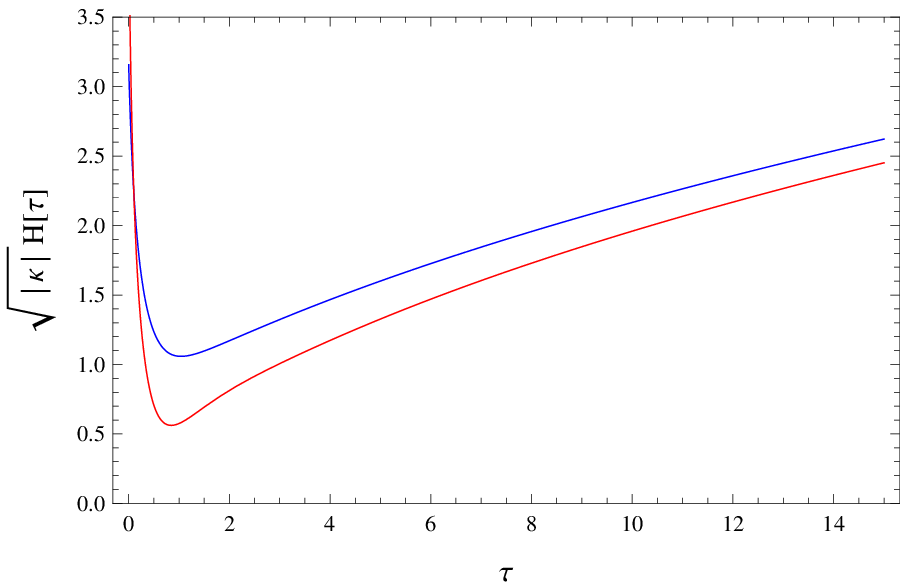}
\end{center}
\end{minipage}
\caption{Examples of the trajectories which approach the asymptote $y = [2/ (3 \pi \gamma V_0)]^{1/2}x^{-3}$ when $\kappa$ is negative (left panel), 
and the corresponding time evolution of the Hubble parameter $H(t)$ (right panel). 
The values of constants are taken as $V_0=0.1$, $\gamma = 0.1$ and $\lambda =0.2$.}
\label{nkt2}
\end{figure}

For negative $\kappa$ there are two possible asymptotical regimes for the scalar field given by Eqs. \Ref{as1m} and \Ref{asphi3m}.
They have the same form as the respective asymptotics \Ref{as1} and \Ref{asphi3} obtained for $\kappa>0$. Therefore, in the case $\kappa<0$
there exist two possible cosmological scenarios of the late-time evolution of the Universe:
(i) the inflationary (quasi-de Sitter) expansion with $H(t)\simeq H_\infty=\sqrt{\frac23\pi G\lambda\phi_0^4}$, and
(ii) the Little Rip scenario given by Eqs. \Ref{LittleRip-phi}, \Ref{LittleRip-H}, \Ref{LittleRip-a}.
These two scenarios are depicted in Fig.~\ref{nkt1} and Fig.~\ref{nkt2}, respectively.


\section{Higgs field as a driver of the accelerated expansion}
\label{Higgs}
Above we have found that the Higgs field $\phi$ with the potential $V(\phi)=\lambda/4(\phi^2-\phi_0^2)^2$ can, in principle, cause an accelerated regime of the Universe evolution.
In the Standard Model the vacuum expectation value is $\phi_0 \simeq 246$ GeV,
and the coupling constant of the Higgs boson is estimated as $\lambda \simeq 0.13$ \cite{Agashe:2014kda}. Now, we can obtain the following estimations:
\begin{eqnarray}
\gamma &=& G\phi_0^2=\frac{\phi_0^2}{M_{pl}^2}\simeq 10^{-34}, \nonumber\\
V_0    &=& 2\pi G\kappa\lambda\phi_0^4 = \left(\frac{2\pi\lambda\phi_0^4}{M_{pl}^2}\right)\kappa
\simeq (3\times 10^{-29}\, \mbox{GeV}^2) \, \kappa,
\label{higgsestimations}
\end{eqnarray}
where $M_{pl}=G^{-1/2}\simeq 10^{19}$ GeV is the Planck mass.

As was shown, in order to realize the late-time de Sitter scenario \Ref{latetimeinf} in case $\kappa>0$,
the condition $V_0=2\pi G\kappa\lambda\phi_0^4>1$ should be fulfilled. Now, using the estimations \Ref{higgsestimations},
we find $\kappa \simeq 3 \times 10^{10}$ eV$^{-2}$, or the corresponding length $l_\kappa=\kappa^{1/2}\simeq 4~\mbox{cm}$.
For comparison, in Refs. \cite{Sus:2009,SarSus:2010,Sus:2012} it was shown that the nonminimal kinetic coupling provides
the early-time inflationary epoch with the Hubble parameter $H_\kappa=1/\sqrt{9\kappa}$, and in \cite{Sus:2012}
the estimations $\kappa\simeq 10^{-74}$ sec$^2$ and $l_\kappa\simeq 10^{-27}$ cm had been given.
The values of $\kappa$ and $l_{\kappa}$ obtained here are of many orders larger.
While, the energy scale of the Hubble rate $H = \sqrt{\frac23\pi G\lambda\phi_0^4}\simeq 10^{-6} ~\text{eV}$ is too small for the initial inflation and
too large for the late-time accelerated expansion of the Universe.

Another scenario of an accelerated expansion is represented by the Big Rip asymptotics
\Ref{BigRip-phi}, \Ref{BigRip-H}, \Ref{BigRip-a}. It could be realized for {\it any} positive $\kappa$.
During the Big Rip expansion, the scalar field is infinitely growing,
so that $\phi\gg\phi_0$ and $V(\phi)\approx \frac{\lambda}{4}\phi^4$.
As the result, a behavior of the scalar field at the Big Rip epoch is only determined by $\kappa$ and $\lambda$ as follows:
$\phi(t)\simeq \sqrt{392\kappa/\lambda}(t_*-t)^{-2}$.
However, it is interesting that the Hubble rate is not depend on $\kappa$ and $\lambda$ at all, i.e. $H^2(t)\simeq(49/9)(t_*-t)^{-2}$.

One more accelerated regime is provided by the Little Rip scenario with the asymptotics \Ref{LittleRip-phi},
\Ref{LittleRip-H}, \Ref{LittleRip-a}.
Note that this scenario is realized for {\it any}, both positive and negative, nonzero $\kappa$.
Since during the Little Rip epoch the scalar field is infinitely growing (see Eq. \Ref{LittleRip-phi}),
we have $\phi\gg\phi_0$ and $V(\phi)\approx \frac{\lambda}{4}\phi^4$ at large times,
and an asymptotical behavior of $\phi(t)$, $H(t)$, and $a(t)$ depends on $\kappa^2$, $\lambda$, and $G$ (or $M_{pl}^2$).


\section{Summary}
\label{Conclusion}
In this paper we have investigated cosmological dynamics of the FRW Universe filled with
the scalar field $\phi$,
which has the potential $V(\phi)=\frac{\lambda}{4}(\phi^2-\phi_0^2)^2$ and possesses the nonminimal kinetic coupling
to the curvature given as $\kappa G^{\mu\nu}\phi_{,\mu}\phi_{,\nu}$, where $\kappa$ is a coupling parameter with dimension of $(length)^2$.
The most important and interesting result obtained is that
the scalar field with the kinetic coupling
provides a new mechanism to generate accelerated regimes of the Universe evolution.
This mechanism is working due to nonminimal kinetic coupling and does not need any phantom matter.
There are three possible cosmological scenarios with an acceleration.

\begin{enumerate}
\item {\it The late-time de Sitter scenario.}
In this scenario the Hubble parameter exponentially tends to the constant value $H_\infty=\sqrt{\frac23\pi G\lambda\phi_0^4}$ at large times,
while the scalar field tends to zero, so that the Higgs potential reaches its local maximum $V(0)=\frac{\lambda}{4}\phi_0^4$.
For positive $\kappa$ this regime is realized only if $2\pi G\kappa\lambda\phi_0^4>1$,
and for negative $\kappa$ the late-time de Sitter scenario could be realized for any parameters of the scalar field.

\item {\it The Big Rip.}
The Big Rip scenario can be realized for any positive $\kappa$.
In this epoch the Hubble parameter is found as follows $H(t)\simeq(49/9)^{1/2}(t_*-t)^{-1}$;
it becomes infinite within a finite interval of time, so that $H(t)\to\infty$ as $t\to t_*$.
The scalar field is also infinitely growing as $\phi(t)\simeq \sqrt{392\kappa/\lambda}(t_*-t)^{-2}$,
so that $\phi\gg\phi_0$ and $V(\phi)\approx \frac{\lambda}{4}\phi^4$.

\item {\it The Little Rip.}
The Little Rip scenario can be realized for any, both positive and negative $\kappa$.
In this epoch the Hubble parameter has been found as $H(t)\simeq (8\lambda/27\pi G\kappa^2)^{1/4}\, t^{1/2}$,
so that $H(t)\propto t^{1/2}\to\infty$ as $t\to\infty$.
The scalar field is also infinitely growing as $\phi(t)\simeq (2/3\pi^3 G^3\lambda\kappa^2)^{1/8}\, t^{1/4}$,
so that $\phi\gg\phi_0$ and $V(\phi)\approx \frac{\lambda}{4}\phi^4$ at $t\to\infty$.
\end{enumerate}

Also, we have derived the modified slow-roll conditions for the Higgs field with the nonminimal kinetic coupling
(see Eqs. \Ref{extracond2}, \Ref{modkinencond3}, \Ref{slowrollcond3}, and \Ref{slowrollcond4}) and shown that,
in contrast to the usual slow-roll conditions which provide an exponential (de Sitter) expansion of the Universe,
the modified conditions lead to the Little Rip scenario.

\section*{Acknowledgments}
We thank Alexei Toporensky for useful discussions. The work was supported by the Russian Government Program of Competitive Growth of Kazan Federal University and, partially, by the Russian Foundation for Basic Research grants No. 14-02-00598 and 15-52-05045.

\end{document}